\documentclass[journal]{IEEEtran}
\ifCLASSINFOpdf
\else
\fi

\usepackage{overpic}
\usepackage{lineno,hyperref}
\usepackage{amssymb}
\usepackage{times}
\usepackage{epsfig}
\usepackage{graphicx}
\usepackage{amsmath}
\usepackage{amssymb}
\usepackage{booktabs}
\usepackage{subeqnarray}
\usepackage{color}
\usepackage{amstext}
\usepackage{array}
\usepackage{subfigure}
\usepackage{setspace}
\usepackage{chngpage}
\usepackage{multirow}
\graphicspath{{figure/},{figure1/},{figure2/},{figure5/},{figure3/},{figure4/},{figure6/},{figure7/}}
\modulolinenumbers[5]

\begin{document}

\title{Progressive Training of Multi-level Wavelet\\ Residual Networks for Image Denoising} 
%
%
%

\author{Yali Peng,~
        Yue~Cao,
        Shigang~Liu,~
        Jian~Yang,~
        and~Wangmeng~Zuo,~\IEEEmembership{Senior~Member,~IEEE}
\thanks{This project is partially supported by the National Natural Science
Foundation of China (NSFC) under Grant No.s 61873155, 61672333 and U19A2073.}
\thanks{Yali Peng, Yue Cao and Shigang Liu are Key Laboratory of Modern Teaching Technology, Ministry of Education, Xi'an 710062, China, and also with the School of Computer Science, Shaanxi Normal University, Xi'an 710119, China, e-mail: pengyl@snnu.edu.cn, cscaoyue@gmail.com, shgliu@snnu.edu.cn.}
\thanks{Jian Yang is with the School of Computer Science and Engineering, Nanjing University of Science and Technology, Nanjing 210094, China. e-mail: csjyang@njust.edu.cn.}
\thanks{Wangmeng Zuo is with the School of Computer Science and Technology, Harbin Institute of Technology, Harbin 150001, China, e-mail: cswmzuo@gmail.com.}
\thanks{The first two authors contributed equally to this work.}
\thanks{Manuscript received xx xx, 2020; revised xx xx, xxxx.}}

%
%

\markboth{Journal of \LaTeX\ Class Files,~Vol.~14, No.~8, August~2015}%
{Shell \MakeLowercase{\textit{et al.}}: Bare Demo of IEEEtran.cls for IEEE Journals}
%



\maketitle

\begin{abstract}
Recent years have witnessed the great success of deep convolutional neural networks (CNNs) in image denoising.
Albeit deeper network and larger model capacity generally benefit performance, it remains a challenging practical issue to train a very deep image denoising network.
Using multi-level wavelet-CNN (MWCNN) as an example, we empirically find that the denoising performance cannot be significantly improved by either increasing wavelet decomposition levels or increasing convolution layers within each level.
To cope with this issue, this paper presents a multi-level wavelet residual network (MWRN) architecture as well as a progressive training (PT-MWRN) scheme to improve image denoising performance.
In contrast to MWCNN, our MWRN introduces several residual blocks after each level of discrete wavelet transform (DWT) and before inverse discrete wavelet transform (IDWT).
For easing the training difficulty, scale-specific loss is applied to each level of MWRN by requiring the intermediate output to approximate the corresponding wavelet subbands of ground-truth clean image.
To ensure the effectiveness of scale-specific loss, we also take the wavelet subbands of noisy image as the input to each scale of the encoder.
Furthermore, progressive training scheme is adopted for better learning of MWRN by beigining with training the lowest level of MWRN and progressively training the upper levels to bring more fine details to denoising results.
Experiments on both synthetic and real-world noisy images show that our PT-MWRN performs favorably against the state-of-the-art denoising methods in terms both quantitative metrics and visual quality.
\end{abstract}


%
\IEEEpeerreviewmaketitle

\section{Introduction}
%
%
%
%
\IEEEPARstart{I}{mage} denoising is a fundamental yet active task in image processing and low level vision with many real-world applications.
On the one hand, a great number of methods~\cite{rudin1992nonlinear,tomasi1998bilateral,elad2006image,dabov2008image} have been developed for handling synthetic noise such as additive white Gaussian noise (AWGN) with known noise variance $\sigma^2$.
With the advent of deep learning, convolutional neural networks (CNNs)~\cite{Zhang2016Beyond,mao2016image,liu2018multi} have shown significant promise for image denoising and remarkably boosted the denoising performance.
On the other hand, real-world image noise is generally much more complex in contrast to synthetic noise.
In the recent few years, several attempts have been made to develop CNN denoisers for handling real-world noisy photographs by either modeling sophisticated real noise~\cite{guo2019toward, brooks2019unprocessing}, acquiring noisy and nearly noise-free image pairs~\cite{plotz2017benchmarking, SIDD_2018_CVPR}, or learning denoising models in an unsupervised or self-supervised manner~\cite{krull2019noise2void,SelfDenoising,DBSN}.

Albeit remarkable progress has been made, it remains a challenging issue to train a high-performance deep denoising network.
One direct solution for boosting denoising performance is to enlarge deep network capacity.
However, simply increasing network depth and width cannot always benefit the performance improvement.
For example, Table~\ref{tab:mwcnn_variants} lists the results of several variants of multi-level wavelet-CNN (MWCNN)~\cite{liu2018multi} by increasing wavelet decomposition levels, convolution layers in each level, and network width.
In comparison with the original MWCNN~\cite{liu2018multi}, the quantitative gains by MWCNN variants with more decomposition levels and more convolutional layers are very limited, i.e., at most 0.03 dB by PSNR on the BSD68~\cite{martin2001database} when $\sigma = 25$.
On the other hand, 0.04 dB gain on BSD68 can be attained by increasing the network width to its 1.5 times, but further increasing network width only brings 0.01 dB gain.
Thus, it remains a challenging issue to train a deep denoising network with high model capacity, and the denoising performance cannot be significantly improved by increasing network depth and width.


In this paper, we develop a deeper and effective network, i.e.,  multi-level wavelet residual network (MWRN), by improving MWCNN in terms of network architecture, loss functions and training scheme.
In MWCNN~\cite{liu2018multi}, four convolutional layers are deployed after each level of wavelet transform (DWT) and before inverse discrete wavelet transform (IDWT).
However, simply stacking more convolutional layers suffers from the performance degradation problem~\cite{He2016Deep}.
That is, along with the increase of network depth, even the training error may stop to decrease or begin to increase.
Following~\cite{He2016Deep}, we substitute the last three convolutional layers in each level (i.e., scale) of MWCNN with several residual blocks (i.e., four in this work), resulting in our MWRN architecture.

The introduction of residual blocks, however, cannot completely address the training difficulty issue.
In~\cite{lee2015dsn}, Lee \emph{et al.} showed that the deployment of intermediate supervision on hidden layers is effective in improving the model transparency and easing training difficulty. %
Considering the multi-scale characteristics of MWRN, it is natural to constrain the intermediate output for each level of MWRN to approximate the corresponding wavelet subbands of ground-truth clean image.
Therefore, we suggest to deploy scale-specific loss to each level of MWRN.
Nonetheless, we empirically find that scale-specific loss benefits moderately to denoising performance when we only take the noisy image as input to the first layer of MWRN.
As a remedy, for each wavelet decomposition level in the encoder, we also take the wavelet subbands of noisy image as input, and apply an extra convolutional layer to combine it with the feature map from the upper scale.
\begin{table}[!t]
\centering
\caption{Average PSNR(dB) results of several variants of MWCNN with the noise level $\sigma$ $=$ $25$.
In particular, the variant ($D$,$C$,$\rho$) involves $D$ decomposition levels, $C$ convolutional layers are deployed after each DWT and before each IDWT operation, and the number of channels is $\rho$ times of that of the default MWCNN~\cite{liu2018multi}.
}
\label{tab:mwcnn_variants}
\scalebox{1}{
\begin{tabular}{cccc}
\toprule[1pt]
($D$, $C$, $\rho$)&
Set12&
BSD68&
Urban100\\
\bottomrule[1pt]
(3,4,1)&
30.79&
29.41&
30.66\\

(4,4,1)&
30.82&
29.42&
30.73\\

(5,4,1)&
30.83&
29.42&
30.74\\

(3,7,1)&
30.84&
29.44&
30.77\\

(3,9,1)&
30.80&
29.42&
30.67\\

(3,4,1.5)&
30.86&
29.45&
30.80\\

(3,4,2)&
30.87&
29.46&
30.83\\
\bottomrule[1pt]
PT-MWRN &
30.96&
29.51&
31.00\\
\bottomrule[1pt]

\end{tabular}
}
\end{table}

We further present a progressive training scheme for effective learning of MWRN, resulting in our PT-MWRN method.
Due to the noisy image is taken as the input for each decomposition level, it is feasible to pre-train the lower levels of MWRN without the upper levels.
Thus, we first train the lowest scale of MWRN and subsequently train the upper scales to bring more fine details to the denoising results.
Progressive training begins with the pre-training of a shallow network.
Then, the upper scale of MWRN can be gradually trained by initializing the lower scales with the pre-trained parameters.
Therefore, progressive training is also beneficial to ease the training of deeper denoising network.
With the introduction of MWRN, scale-specific loss and progressive training, PT-MWRN can notably boost the denoising performance of MWCNN.
It can be seen from Table~\ref{tab:mwcnn_variants} that PT-MWRN achieve PSNR gains of 0.17/0.10/0.34 dB against MWCNN for removing AWGN with $\sigma = 25$ respectively on the Set12/BSD68/Urban100 datasets.

Extensive experiments are conducted to evaluate our PT-MWRN on both grayscale and color image denoising.
Ablation studies show that both residual blocks, scale-specific loss and progressive training can be utilized to ease the training of MWRN, and are effective in improving denoising performance.
For Gaussian denoising, PT-MWRN performs favorably against the state-of-the-art denoising methods, e.g., BM3D \cite{dabov2008image}, TNRD \cite{chen2017trainable}, DnCNN \cite{Zhang2016Beyond}, IRCNN \cite{zhang2017learning}, RED \cite{mao2016image}, MemNet \cite{tai2017memnet}, N3Net \cite{Ploetz:2018:NNN}, NLRN \cite{liu2018non} and MWCNN \cite{liu2018multi} for grayscale images, and  CBM3D \cite{dabov2008image}, IRCNN \cite{zhang2017learning}, CDnCNN \cite{Zhang2016Beyond}, FFDNet \cite{zhang2018ffdnet}, DHDN \cite{park2019densely} and CMWCNN \cite{liu2018multi} for color images.
When paired real-world noisy and clean images are available, PT-MWRN can be easily extended to handle real-world noisy images.
The results on Darmstadt Noise Dataset (DND) \cite{plotz2017benchmarking} and Smartphone Image Denoising Dataset (SIDD) \cite{SIDD_2018_CVPR} show that our PT-MWRN achieves the state-of-the-art denoising performance on real-world noisy photographs.
Generally, the main contribution of this paper can be summarized as follows:
\begin {itemize}
   \item A deeper multi-level wavelet residual network (MWRN) is presented for effective image denoising.
       In comparison to MWCNN, residual blocks are introduced in each level of the encoder and decoder, and the wavelet subbands of noisy image are taken as input for each decomposition level in the encoder.
       %

    \item Scale-specific loss and progressive training are further introduced to ease the training of MWRN and improve denoising performance, resulting in our PT-MWRN.

    \item Quantitative and qualitative results clearly show that our PT-MWRN performs favorably against the state-of-the-art methods for both Gaussian denoising on grayscale and color images and for handling real-world noisy photographs.

%
%
 \end {itemize}

%
%
%
%

The remainder of this paper is organized as follows.
Section~\ref{sec:Related Work} briefly surveys the progress in deep denoising models for handling AWGN and real-world noise.
Section~\ref{sec:PROPOSED METHOD} presents the network architecture, scale-specific loss and progressive training scheme.
Section~\ref{sec:Experiments} reports the experimental results for images with both AWGN and real-wrold noise.
Finally, Section~\ref{sec:CONCLUSION} ends this work with several concluding remarks.


\section{Related Work}
In this section, we present a brief review on the development of deep denoising networks for removing AWGN and real-world image noise.
\label{sec:Related Work}

\subsection{Learning-based Gaussian Denoising}
As a fundamental image processing task, image denoising has been continuously and intensively studied for decades.
Many traditional methods, e.g., total variation (TV)~\cite{rudin1992nonlinear}, bilateral filtering~\cite{tomasi1998bilateral}, sparse representation \cite{elad2006image}, fields of experts (FoE)~\cite{roth2005fields}, and nonlocal self-similarity (NSS)~\cite{dabov2008image,gu2014wnnm}, have been proposed.
Among them, BM3D~\cite{dabov2008image} and WNNM~\cite{gu2014wnnm} are two representative NSS-based methods with decent denoising performance.
Model-guided discriminative learning has also been introduced to image denoising by learning the image prior model parameters as well as compact unrolled inference in a discriminative manner~\cite{schmidt2014csf,chen2017trainable,lefkimmiatis2017cvpr}.
For example, Schmidt \emph{et al.}~\cite{schmidt2014csf} presented a cascade of
shrinkage fields (CSF) model by incorporating half quadratic splitting (HQS) optimization with the FoE model.
Furthermore, Chen \emph{et al.}~\cite{chen2017trainable} suggested a trainable nonlinear reaction diffusion (TNRD) approach to learn truncated gradient descent inference.
And Lefkimmiatis~\cite{lefkimmiatis2017cvpr} incorporated the NSS-based image prior with the learning of proximal gradient-based inference.

%
%
%
%

Instead of discriminative learning on image priors and unrolled inference, deep denoising networks aim at learning a direct mapping from the noisy image to the desired clean image.
Benefitted from the rapid progress in deep learning, convolutional neural networks (CNNs) have achieved very competitive performance in comparison to both traditional methods and model-guided discriminative learning~\cite{Zhang2016Beyond,mao2016image}.
In particular, Zhang \emph{et al.}~\cite{Zhang2016Beyond} analyzed the connection between model-guided discriminative learning and deep CNNs, and incorporated residual learning~\cite{He2016Deep} with batch normalization (BN)~\cite{ioffe2015batch} to constitute a denoising convolutional network (DnCNN).
While Mao \emph{et al.}~\cite{mao2016image} presented a deep fully convolutional encoding-decoding network with symmetric skip connections (i.e., RED).

Driven by the advances in deep CNNs, more and more network architectures (e.g., U-Net~\cite{ronneberger2015u}, ResNet~\cite{He2016Deep}, and DenseNet~\cite{huang2017densely}) and modules have been introduced to the design of deep denoising networks and continually improved the Gaussian denoising performance.
In~\cite{zhang2019rdn}, dense connected layers and local residual learning are incorporated to form residual dense block (RDB) for building the residual dense network (RDN).
By combining residual on the residual structure and channel attention, Anwar and Barnes~\cite{anwar2019real} presented a real image denoising network (RIDNet).
Following the non-local neural networks~\cite{wang2018nonlocal}, non-local module has also been introduced to deep denoising networks.
In particular, NLRN~\cite{liu2018non} incorporates non-local module with recurrent networks, while N3Net~\cite{Ploetz:2018:NNN} relaxes the k-nearest neighbors (KNN) selection with neural nearest neighbors block.

%
%
%
%

Several recent studies have also been given to incorporate wavelet transform~\cite{Mallat1996Wavelets} with CNNs for boosting denoising performance.
Bae \emph{et al.} \cite{bae2017beyond} presented a wavelet residual network to predict the clean image in wavelet domain.
Liu \emph{et al.}~\cite{liu2018multi} suggested a multi-level wavelet CNN (MWCNN), where DWT and IDWT are respectively deployed to substitute downsampling in encoder and up-convolution in decoder.
Besides, wavelet transform has also been exploited in building deep CNNs for other low level vision tasks such as single image super-resolution~\cite{guo2017deep,huang2017wavelet,deng2019iccv}.
Albeit state-of-the-art denoising performance has been achieved, it remains a challenging issue to train a deeper wavelet-based denoising network (e.g., MWCNN).
Therefore, we present a deeper multi-level wavelet residual network (MWRN), and incorporate scale-specific loss and progressive training to ease the model training.
Moreover, albeit our PT-MWRN is suggested for improving MWCNN, the idea of scale-specific loss
and progressive training may also be beneficial to the training of other multi-scale CNNs in low level vision.

\subsection{Deep Networks for Real-world Noisy Photographs}
For real-world noisy photographs, the noise model is more sophisticated than AWGN and the parameters of noise model is unknown.
Consequently, deep networks for blind Gaussian denoising generally perform poorly on real-world noisy images~\cite{plotz2017benchmarking}.
One plausible solution is to develop realistic noise model, which is then exploited to synthesize noisy images from clean images and to train deep denoisers.
Guo~\emph{et al.}~\cite{guo2019toward} presented a realistic noise model by considering heteroscedastic Gaussian noise and in-camera signal processing (ISP) of raw pixels, and developed a convolutional blind denoising network (CBDNet).
Brooks \emph{et al.}~\cite{brooks2019unprocessing} also adopted signal-dependent heteroscedastic Gaussian for modeling raw image noise, and suggested to first unprocess sRGB image to raw measurement and then convert the denoised raw image to sRGB space.
Wei \emph{et al.}~\cite{wei2020physical} provided an accurate raw image noise model for CMOS sensor by considering the imaging pipeline from photons to electrons, from electrons to voltage, and from voltage to digital numbers.
Instead of explicit noise modeling, learning based models have also been introduced for modeling real-world noise and ISP procedure~\cite{abdelhamed2019noiseflow,Zamir2020CycleISP}.

%
%

Besides, several approaches have been developed to acquire the nearly noise-free image of given real-world noisy photograph.
In particular, Pl\"{o}tz and Roth~\cite{plotz2017benchmarking} aligned and post-processed the low ISO image to obtain nearly noise-free image, while \cite{SIDD_2018_CVPR,nam2016holistic} achieved this goal by aligning and averaging multiple noisy images.
Based on the pairs of synthetic noisy-clean images and the pairs of real noisy and nearly noise-free images, deep denoising networks~\cite{guo2019toward,yue2019variational,anwar2019real} can be trained for handling real-world noisy photographs.
Path-Restore~\cite{yu2019path} adopted a multi-path CNN together with a pathfinder for dynamically selecting appropriate route.
In NTIRE 2019 Challenge on Real Image Denoising~\cite{abdelhamed2019ntire}, GRDN\cite{kim2019grdn}, DHDN~\cite{park2019densely} and DIDN~\cite{yu2019deep} have won the first three places on the sRGB track.
GRDN~\cite{kim2019grdn} incorporated grouped residual dense blocks (GRDBs) and convolutional block attention module (CBAM)~\cite{woo2018cbam}, while  DHDN \cite{park2019densely} and DIDN \cite{yu2019deep} are based on the modified U-Net \cite{ronneberger2015u} architectures.
In comparison to the above methods, our PT-MWRN trained using SIDD training set and synthetic images can achieve very competitive denoising results on the DND and SIDD benchmarks.

%


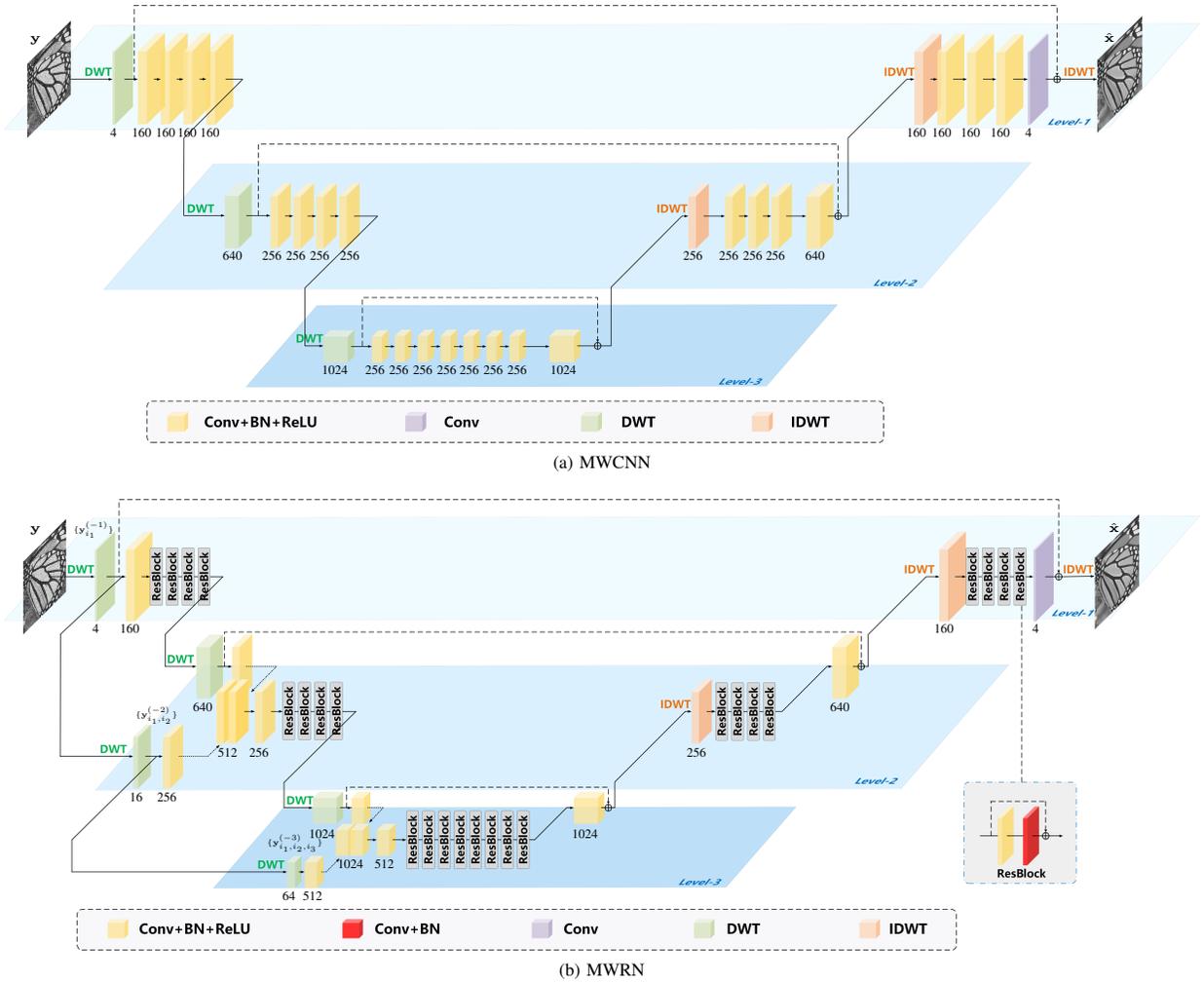
\begin{figure*}[!htbp]
\scriptsize{
\begin{center}
\vspace{-0ex}

\subfigure{
\centering

\begin{overpic}[width=0.9\textwidth,scale=0.5]{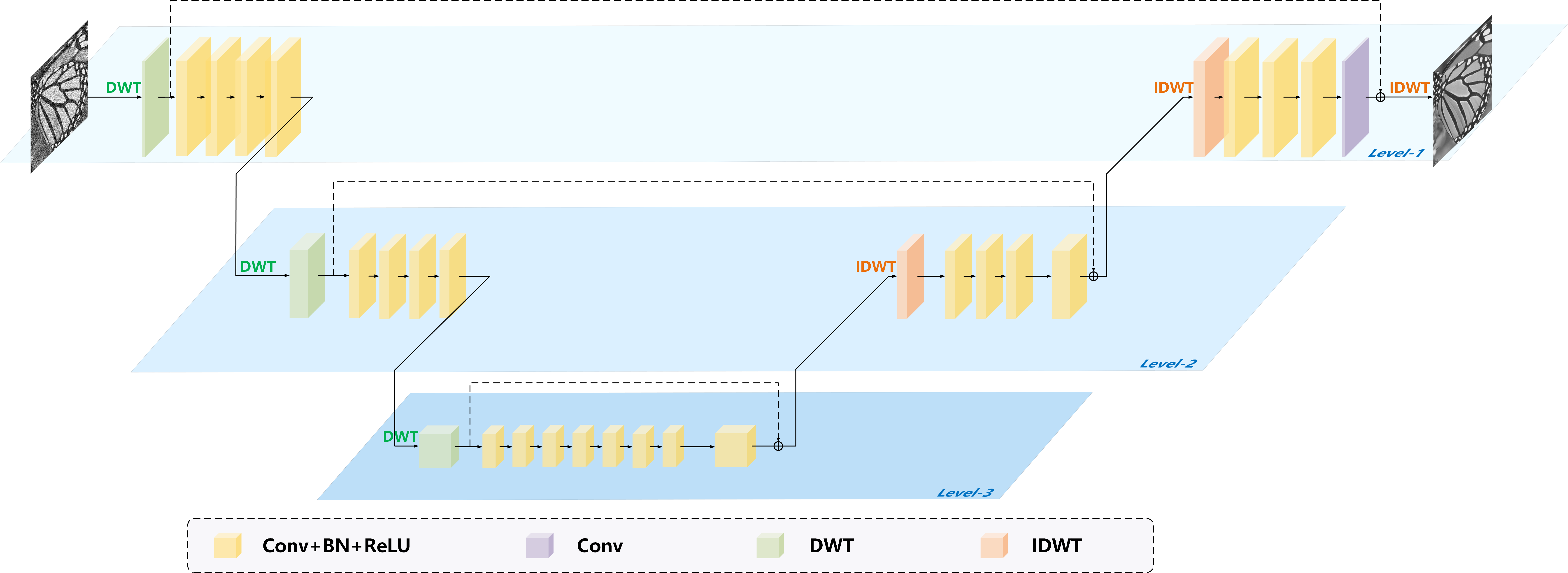} 
\put(2.2,33.5){\color{black}\scalebox{1}{{\tiny$\mathbf{y}$}}}

\put(92,33.5){\color{black}\scalebox{1}{{\tiny$\hat{\mathbf{x}}$}}}

\put(8.9,25.6){\color{black}{\tiny 4}}
\put(10.7,25.6){\color{black}{\tiny 160}}
\put(12.7,25.6){\color{black}{\tiny 160}}
\put(14.5,25.6){\color{black}{\tiny 160}}
\put(16.4,25.6){\color{black}{\tiny 160}}
\put(18.3,15.3){\color{black}{\tiny 640}}
\put(21.6,15.3){\color{black}{\tiny 256}}
\put(23.6,15.3){\color{black}{\tiny 256}}
\put(25.6,15.3){\color{black}{\tiny 256}}
\put(28.1,15.3){\color{black}{\tiny 256}}
\put(26.6,5.7){\color{black}{\tiny 1024}}
\put(30.2,5.7){\color{black}{\tiny 256}}
\put(32.2,5.7){\color{black}{\tiny 256}}
\put(34.2,5.7){\color{black}{\tiny 256}}
\put(36.2,5.7){\color{black}{\tiny 256}}
\put(38.1,5.7){\color{black}{\tiny 256}}
\put(40.1,5.7){\color{black}{\tiny 256}}
\put(42.1,5.7){\color{black}{\tiny 256}}
\put(45.7,5.7){\color{black}{\tiny 1024}}
\put(56.8,15.3){\color{black}{\tiny 256}}
\put(59.8,15.3){\color{black}{\tiny 256}}
\put(61.8,15.3){\color{black}{\tiny 256}}
\put(63.7,15.3){\color{black}{\tiny 256}}
\put(67,15.3){\color{black}{\tiny 640}}
\put(75.5,25.6){\color{black}{\tiny 160}}
\put(77.6,25.6){\color{black}{\tiny 160}}
\put(80,25.6){\color{black}{\tiny 160}}
\put(82.5,25.6){\color{black}{\tiny 160}}
\put(85.4,25.6){\color{black}{\tiny 4}}
\end{overpic}

  \label{fig:mwcnn}
}\\
\centerline{\scriptsize (a) MWCNN}
\vspace{+3ex}
\subfigure{

\centering

 \begin{overpic}[width=0.9\textwidth,scale=0.5]{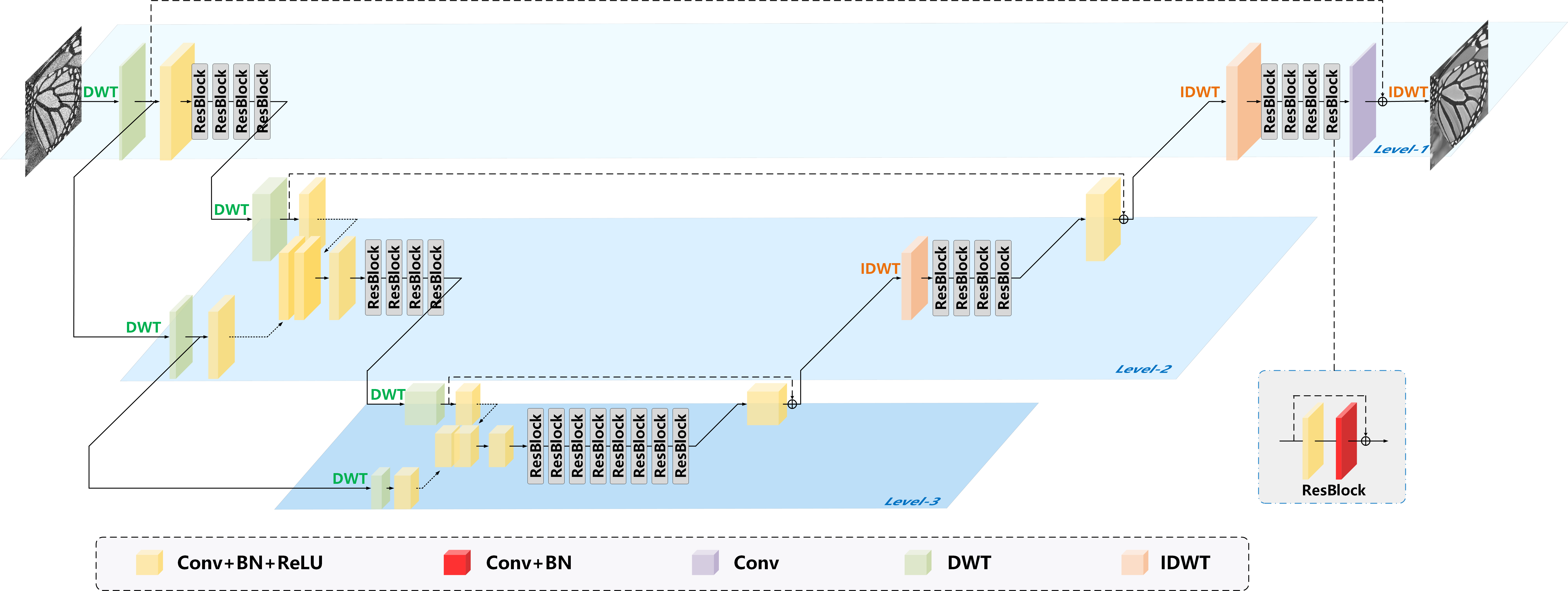} 

\put(22,8.8){\color{black}\scalebox{0.6}{{\tiny$\{ \mathbf{y}_{i_1, i_2, i_3}^{(-3)} \}$}}}

\put(11,19.5){\color{black}\scalebox{0.6}{{\tiny$\{\mathbf{y}_{i_1, i_2}^{(-2)}\}$}}}

\put(5.8,35){\color{black}\scalebox{0.6}{{\tiny$\{\mathbf{y}_{i_1}^{(-1)}\}$}}}

\put(2.2,35){\color{black}\scalebox{1}{{\tiny$\mathbf{y}$}}}

\put(92.4,35){\color{black}\scalebox{1}{{\tiny$\hat{\mathbf{x}}$}}}

\put(7.4,26.6){\color{black}{\tiny 4}}
\put(9.7,26.6){\color{black}{\tiny 160}}
\put(15.8,20){\color{black}{\tiny 640}}
\put(10.5,12.5){\color{black}{\tiny 16}}
\put(12.8,12.5){\color{black}{\tiny 256}}
\put(17.9,16.2){\color{black}{\tiny 512}}
\put(20.5,16.2){\color{black}{\tiny 256}}
\put(25.5,9.4){\color{black}{\tiny 1024}}
\put(23.3,4.2){\color{black}{\tiny 64}}
\put(25,4.2){\color{black}{\tiny 512}}
\put(27.9,6.9){\color{black}{\tiny 1024}}
\put(31,6.9){\color{black}{\tiny 512}}
\put(47.5,9.4){\color{black}{\tiny 1024}}
\put(57.1,16.2){\color{black}{\tiny 256}}
\put(69.1,20){\color{black}{\tiny 640}}
\put(77.8,26.6){\color{black}{\tiny 160}}
\put(86,26.6){\color{black}{\tiny 4}}

\end{overpic}

  \label{fig:mwrn}
}
\centerline{\scriptsize (b) MWRN}
\vspace{-2ex}
\caption{\small
{Network architectures of MWCNN and our MWRN.}}
\label{fig:mwtwo}
\vspace{-4ex}
\end{center}}
\end{figure*}

\section{Proposed Method}
In this section, we present a multi-level wavelet residual network (MWRN) and a progressive training scheme for improving denoising performance.
To begin with, we briefly discuss the training difficulty issue of deeper MWCNN.
MWRN is then provided by introducing residual blocks and adding input layer to each scale of MWCNN.
Furthermore, scale-specific loss and progressive training are suggested to ease the training of MWRN, resulting in our PT-MWRN.
\label{sec:PROPOSED METHOD}

\subsection{Training Difficulty of Deeper MWCNN}
\label{sec:mwcnn}
Wavelet transform~\cite{Mallat1996Wavelets} provides an effective signal decomposition with attractive time-frequency localization characteristics, while inverse wavelet transform can accurately reconstruct the original signal for the wavelet subbands.
Motivated by these characteristics, MWCNN adopts the U-Net architecture, and utilizes DWT and IDWT to respectively replace the downsampling operation in encoder and the up-convolution operation in decoder.
Owing to the appealing time-frequency localization of wavelet transform, MWCNN has achieved state-of-the-art denoising performance and is effective in recovering fine image details and structures.
Fig.~\ref{fig:mwcnn} shows the network structure of MWCNN.
In general, it involves $D$ DWT and $D$ IDWT operations.
After each DWT and before each IDWT operation, $C$ convolutional layers are deployed.
Specifically, we have the number of convolutional layers $C = 4$ and the number of scales $D = 3$ for the default MWCNN in Fig.~\ref{fig:mwcnn}, and the number of channels in each layer is also shown in the figure.

%

To further improve the denoising performance, one direct solution is to enlarge the model capacity  of MWCNN, e.g., increasing the $C$ and $D$ values, or the number of channels.
Denote by MWCNN($D$, $C$, $\rho$) an implementation of MWCNN with the number of convolutional layers $C$, the number of scales $D$.
Moreover, the number of channels of MWCNN($D$, $C$, $\rho$) is  $\rho$ times of that of the default MWCNN in Fig.~{\ref{fig:mwcnn}}.
From Table~\ref{tab:mwcnn_variants}, it can be seen that the increase of $D$ and $C$ values benefits little to denoising performance gain.
While 0.04 dB gain on BSD68 can be attained by increasing $\rho$ from 1 to $1.5$, further
increasing $\rho$ to 2 only brings $0.01$ dB gain.
The results indicate that it remains a challenging issue to train deeper/wider MWCNN.
One possible cause to training difficulty is gradient vanishing/exploding.
In the following, we will alleviate this issue from the aspects of network architecture, loss functions, and training strategy.

\subsection{Network Architecture of MWRN}
\label{sec:mwrn}

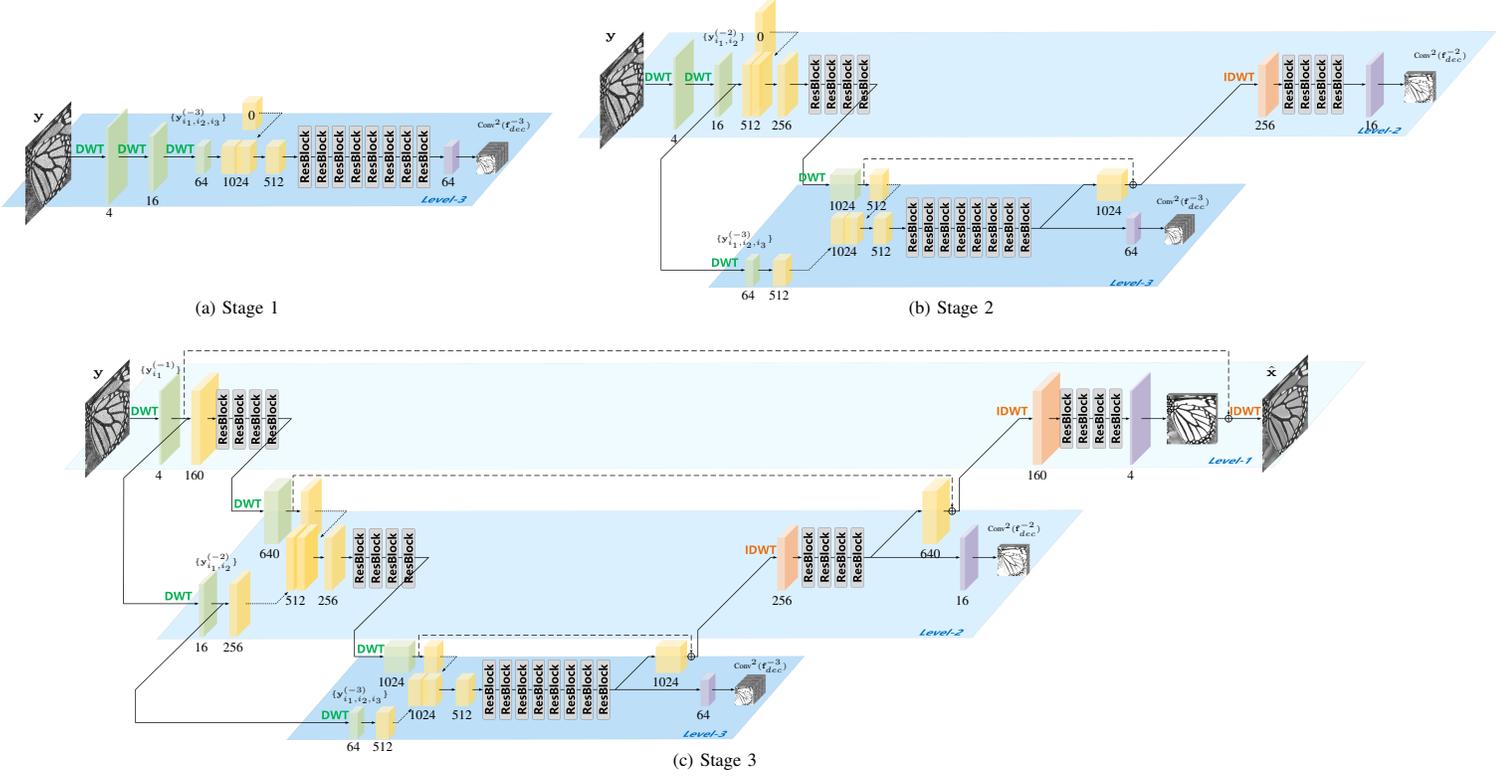
\begin{figure*}[!htbp]
\scriptsize{

\vspace{-4ex}
\hspace{-5ex}
\subfigure{

\begin{overpic}[width=0.4035\textwidth]{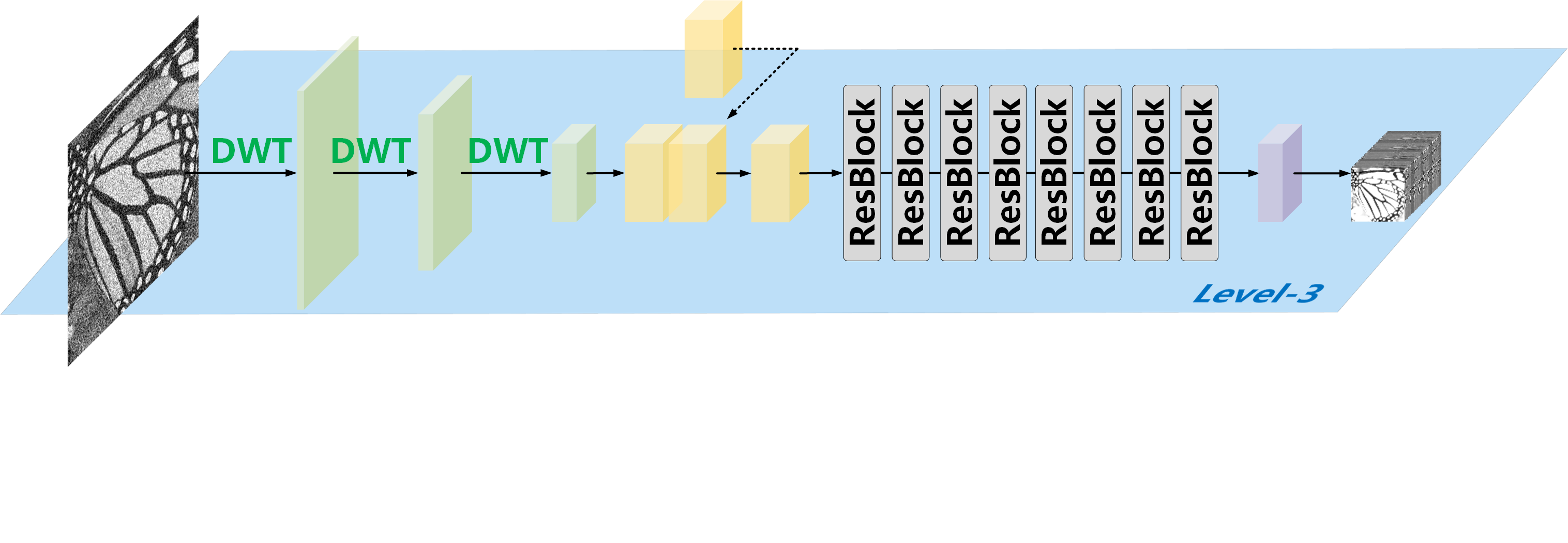} 
\put(86.5,29){\color{black}\scalebox{0.6}{{\tiny $\mbox{Conv}^2(\mathbf{f}^{-3}_{dec})$}}}

\put(30.5,30.5){\color{black}\scalebox{0.6}{{\tiny$\{ \mathbf{y}_{i_1, i_2, i_3}^{(-3)} \}$}}}

\put(5.7,30.5){\color{black}\scalebox{1}{{\tiny$\mathbf{y}$}}}

\put(18.9,12.8){\color{black}{\tiny 4}}
\put(26.1,15){\color{black}{\tiny 16}}
\put(35.1,18.3){\color{black}{\tiny 64}}
\put(40.1,18.3){\color{black}{\tiny 1024}}
\put(47.6,18.3){\color{black}{\tiny 512}}
\put(79.9,18.3){\color{black}{\tiny 64}}
\put(44.8,30.5){\color{black}{\tiny 0}}
\end{overpic}

\label{fig:step1}

}
{\scriptsize }
\hspace{-0ex}
\subfigure{
\begin{overpic}[width=0.6735\textwidth]{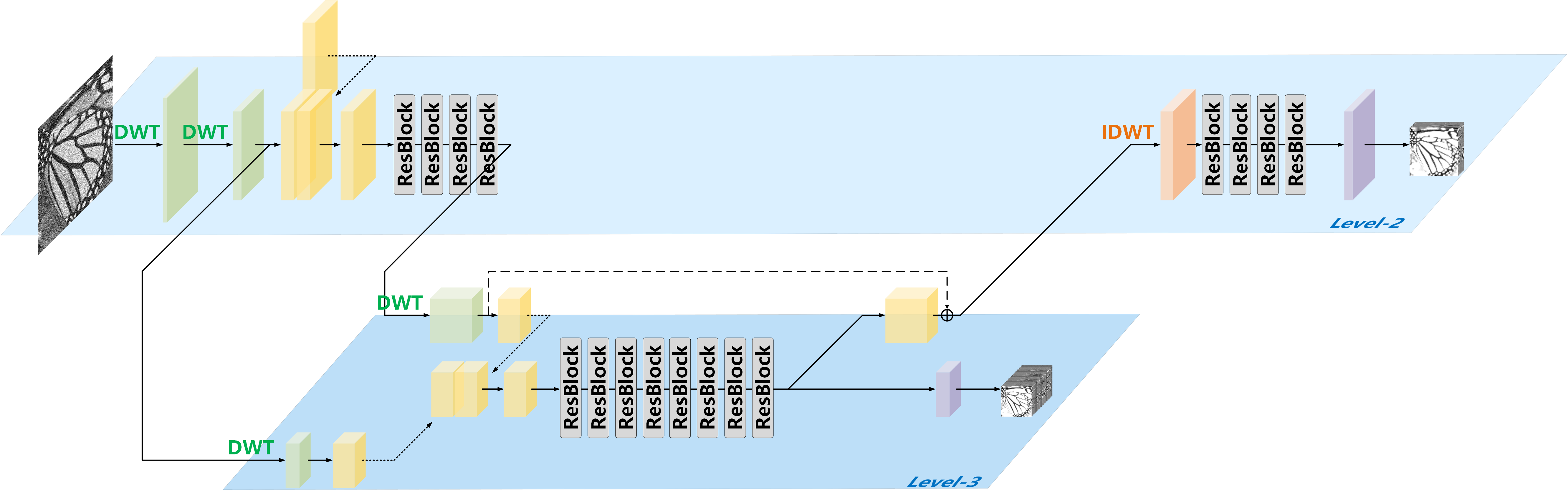} 

\put(3,27){\color{black}\scalebox{1}{{\tiny$\mathbf{y}$}}}
\put(63,9){\color{black}\scalebox{0.6}{{\tiny $\mbox{Conv}^2(\mathbf{f}^{-3}_{dec})$}}}

\put(15,5){\color{black}\scalebox{0.6}{{\tiny$\{ \mathbf{y}_{i_1, i_2, i_3}^{(-3)} \}$}}}

\put(13.5,27){\color{black}\scalebox{0.6}{{\tiny$\{\mathbf{y}_{i_1, i_2}^{(-2)}\}$}}}

\put(91,25){\color{black}\scalebox{0.6}{{\tiny$\mbox{Conv}^2(\mathbf{f}^{-2}_{dec})$}}}

\put(10.1,15.7){\color{black}{\tiny 4}}
\put(14.4,17.2){\color{black}{\tiny 16}}
\put(17.7,17.2){\color{black}{\tiny 512}}
\put(21.1,17.2){\color{black}{\tiny 256}}
\put(73.7,17.2){\color{black}{\tiny 256}}
\put(85.6,17.2){\color{black}{\tiny 16}}
\put(27.3,8.4){\color{black}{\tiny 1024}}
\put(31.4,8.4){\color{black}{\tiny 512}}
\put(17.8,-1.3){\color{black}{\tiny 64}}
\put(20.8,-1.3){\color{black}{\tiny 512}}
\put(27.5,3.4){\color{black}{\tiny 1024}}
\put(31.9,3.4){\color{black}{\tiny 512}}
\put(56.3,7.9){\color{black}{\tiny 1024}}
\put(59.5,3.4){\color{black}{\tiny 64}}
\put(19.5,26.8){\color{black}{\tiny 0}}

\end{overpic}

\label{fig:step2}
}
\centerline{\scriptsize \hspace{-3.2cm}(a) Stage 1 \hspace{+8.2cm} (b) Stage 2}
\begin{center}
\vspace{+2ex}
\subfigure{

\begin{overpic}[width=0.9535\textwidth]{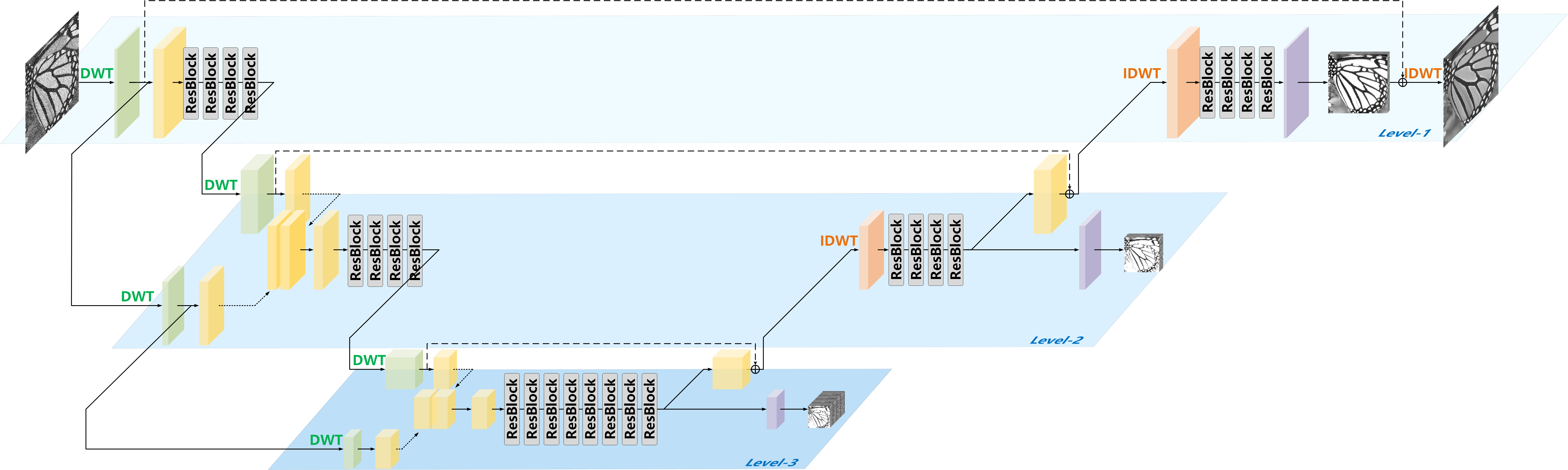} 
\put(51.5,5.5){\color{black}\scalebox{0.6}{{\tiny $\mbox{Conv}^2(\mathbf{f}^{-3}_{dec})$}}}

\put(20.5,3.3){\color{black}\scalebox{0.6}{{\tiny$\{ \mathbf{y}_{i_1, i_2, i_3}^{(-3)} \}$}}}

\put(10,13.5){\color{black}\scalebox{0.6}{{\tiny$\{\mathbf{y}_{i_1, i_2}^{(-2)}\}$}}}

\put(71,16){\color{black}\scalebox{0.6}{{\tiny$\mbox{Conv}^2(\mathbf{f}^{-2}_{dec})$}}}

\put(5.8,28.3){\color{black}\scalebox{0.6}{{\tiny$\{\mathbf{y}_{i_1}^{(-1)}\}$}}}

\put(2.2,28){\color{black}\scalebox{1}{{\tiny$\mathbf{y}$}}}

\put(92.4,28){\color{black}\scalebox{1}{{\tiny$\hat{\mathbf{x}}$}}}

\put(7,20){\color{black}{\tiny 4}}
\put(9.2,20){\color{black}{\tiny 160}}
\put(74,20){\color{black}{\tiny 160}}
\put(81.7,20){\color{black}{\tiny 4}}
\put(15,14){\color{black}{\tiny 640}}
\put(10,6.8){\color{black}{\tiny 16}}
\put(12.2,6.8){\color{black}{\tiny 256}}
\put(17,10.4){\color{black}{\tiny 512}}
\put(19.5,10.4){\color{black}{\tiny 256}}
\put(54.4,10.4){\color{black}{\tiny 256}}
\put(65.8,14){\color{black}{\tiny 640}}
\put(68.5,10.4){\color{black}{\tiny 16}}
\put(24.1,4.1){\color{black}{\tiny 1024}}
\put(45.2,4.1){\color{black}{\tiny 1024}}
\put(21.7,-0.9){\color{black}{\tiny 64}}
\put(23.7,-0.9){\color{black}{\tiny 512}}
\put(26.5,1.6){\color{black}{\tiny 1024}}
\put(29.8,1.6){\color{black}{\tiny 512}}
\put(48.6,1.6){\color{black}{\tiny 64}}
\end{overpic}

\label{fig:step3}
}
\centerline{\scriptsize (c) Stage 3}
\vspace{-2ex}
\caption{\small
{Illustration of the network architectures adopted in the three stages of progressive training.}}
\label{fig:PT}
\vspace{-4ex}
\end{center}}
\end{figure*}

In this subsection, we describe the structure of multi-level wavelet residual network (MWRN) by introducing residual blocks and adding more input layers to MWCNN.
Fig.~\ref{fig:mwrn} illustrates the network architecture of MWRN.
For each scale of MWCNN($D$, $C$, $\rho$), we replace the last $C-1$ convolutional layers in encoder and the first $C-1$ convolutional layers in decoder with $C$ residual blocks.
Using the default MWCNN($3$, $4$, $1$) as an example, the network depth of MWCNN is 24 convolutional layers, while that of MWRN is 58 layers.
As explained in~\cite{He2016Deep}, the deployment of residual blocks is helpful to the gradient back-propagation in training, and is effective in alleviating the gradient vanishing while improving the convergence behaviors.
Thus, albeit MWRN is much deeper than MWCNN, MWRN can be easier to train due to the introduction of residual blocks.

Second, we further modify MWRN by adding an input layer to each scale of the encoder.
Denote by $\mathbf{y}$ a noisy observation with the image size $h \times w$.
In terms of multi-level wavelet packet transform~\cite{1998Ten}, the decomposition result at the $2^{-1}$ scale can be written as,
\begin{equation}\label{eq:dwt}
\{\mathbf{y}_1^{(-1)}, \mathbf{y}_2^{(-1)}, \mathbf{y}_3^{(-1)}, \mathbf{y}_4^{(-1)}\} = DWT(\mathbf{y}),
\end{equation}
where $\mathbf{y}_1^{(-1)}$, $\mathbf{y}_2^{(-1)}$, $\mathbf{y}_3^{(-1)}$, $\mathbf{y}_4^{(-1)}$ are the four wavelet subbands of $\mathbf{y}$.
Subsequently, each subband image $\mathbf{y}_{i_1}^{(-1)}$ (${i_1} = 1, 2, 3, 4$) can be further decomposed into four subbands, i.e., $\mathbf{y}_{i_1,1}^{(-2)}$, $\mathbf{y}_{i_1,2}^{(-2)}$, $\mathbf{y}_{i_1,3}^{(-2)}$, $\mathbf{y}_{i_1,4}^{(-2)}$.
So we can obtain the 16 wavelet subbands $\{\mathbf{y}_{i_1, i_2}^{(-2)} | {i_1} = 1, 2, 3, 4; {i_2} = 1, 2, 3, 4\}$) at the $2^{-2}$ scale.
Analogously, the 64 wavelet subbands $\{\mathbf{y}_{i_1, i_2, i_3}^{(-3)} | {i_1} = 1, 2, 3, 4; {i_2} = 1, 2, 3, 4; {i_3} = 1, 2, 3, 4\}$) at the $2^{-3}$ scale can also be attained.
Then, we take $\{\mathbf{y}_{i_1, i_2}^{(-2)}\}$ as the input to the $2^{-2}$ scale of MWRN.
As illustrated in Fig.~\ref{fig:mwrn}, a $3 \times 3$ convolutional layer is deployed on $\{\mathbf{y}_{i_1, i_2}^{(-2)}\}$ to generate a feature map $\mathbf{F}_{\mathbf{y}}^{-2}$ of 256 channels.
Denote by Conv(DWT($\mathbf{f}^{-1}_{{\mathbf{y}}}$)) the wavelet transform of the first level representation of MWRN encoder.
$\mathbf{F}_{\mathbf{y}}^{-2}$ and Conv(DWT($\mathbf{f}^{-1}_{{\mathbf{y}}}$)) are then concatenated as the input to the next convolutional layer.
Analogously, we also take $\{ \mathbf{y}_{i_1, i_2, i_3}^{(-3)} \}$ as the input to the $2^{-3}$ scale, generate $\mathbf{F}_{\mathbf{y}}^{-3}$ of 512 channels and concatenate it with the wavelet transform Conv(DWT($\mathbf{f}^{-2}_{{\mathbf{y}}}$)) of the second level representation of MWRN encoder.
%

\subsection{Scale-specific Loss}
\label{sec:scale_loss}
The $\ell_2$-norm reconstruction loss in MWCNN is also adopted to train our MWRN,
\begin{equation}\label{eq:l2loss}
\mathcal{L}_r = \frac{1}{2} \| \hat{\mathbf{x}} - \mathbf{x} \|^2,
\end{equation}
where $\hat{\mathbf{x}}$ is the denoising result by MWRN and $\mathbf{x}$ is the ground-truth clean image.

In order to ease the training MWRN, we further introduce scale-specific losses respectively for the $2^{-2}$ and $2^{-3}$ scales.
Let $\mathbf{f}^{-2}_{dec}$ be the feature representation of the $2^{-2}$ scale decoder.
We use IDWT to upsample Conv$^1(\mathbf{f}^{-2}_{dec}$) and combine it with the first level representation $\mathbf{f}^{-1}_{{\mathbf{y}}}$ of MWRN encoder.
Besides, another convolutional layer is also deployed on $\mathbf{f}^{-2}_{dec}$ to predict the wavelet subbands $\{\mathbf{x}_{i_1, i_2}^{(-2)}\}$ at the $2^{-2}$ scale.
Taking the residual learning formulation, we define the scale-specific loss for the $2^{-2}$ scale as follows,
\begin{equation}\label{eq:scaleloss}
\mathcal{L}^s_{-2} = \frac{1}{2} \sum_{i_1} \sum_{i_2} \| \left(\mbox{Conv}^2(\mathbf{f}^{-2}_{dec})\right)_{i_1, i_2} + \mathbf{y}_{i_1, i_2}^{(-2)} - \mathbf{x}_{i_1, i_2}^{(-2)} \|^2.
\end{equation}
Analogously, we can define the scale-specific loss $\mathcal{L}^s_{-3}$ for the $2^{-3}$ scale.
We note that scale-specific losses can serve as intermediate supervision on hidden layers, and thus is effective in improving the model transparency and easing training difficulty.
Moreover, it is empirically found that scale-wise input of wavelet subbands is beneficial to the denoising performance, which may be explained from two aspects.
On the one hand, the successive convolutional layers in the encoder may cause potential information loss, making it difficult to accurately recover the wavelet subbands of lower scales.
On the other hand, the introduction of scale-wise input makes it feasible to utilize the residual learning formulation, and is helpful in improving the convergence behavior.

Finally, the reconstruction and scale-specific losses are combined to constitute our learning objective,
\begin{equation}\label{eq:objective}
\mathcal{L} = \mathcal{L}_r + \lambda (\mathcal{L}^s_{-2} + \mathcal{L}^s_{-3}),
\end{equation}
where $\lambda$ is a tradeoff parameter.

\subsection{Progressive Training}
\label{sec:progressive}
The multi-scale architecture and scale-specific loss also make it feasible to train MWRN in a progressive manner.
In particular, we begin with training the lowest scale of MWRN, and then progressively train the upper scales of MWRN, resulting in our PT-MWRN.
For the lowest scale of MWRN, we only require to train a 19-layer CNN with eight residual blocks.
When training the upper scale of MWRN, the pre-trained lower scales of the network can serve as a good initialization for easing the network training.
In the following, we describe the three stages of the progressive training scheme in more details.

%


%

In the first stage, we take $\{ \mathbf{y}_{i_1, i_2, i_3}^{(-3)} \}$ as the input, and use the scale-specific loss $\mathcal{L}^s_{-3}$ to train the $2^{-3}$ scale of MWRN.
As illustrated in Fig.~\ref{fig:step1}, a $3 \times 3$ convolutional layer is deployed on $\{ \mathbf{y}_{i_1, i_2, i_3} \}$ to generate a feature map $\mathbf{F}_{\mathbf{y}}^{-3}$ of 512 channels.
Due to the upper scales are not trained yet, we simply let the second level feature representation of MWRN encoder, i.e., Conv(DWT($\mathbf{f}^{-2}_{{\mathbf{y}}}$)), be zero.
Then, $\mathbf{F}_{\mathbf{y}}^{-3}$ and Conv(DWT($\mathbf{f}^{-2}_{{\mathbf{y}}}$)) are concatenated and fed into the next convolutional layer.
After eight residual blocks, we can obtain the feature representation  $\mathbf{f}^{-3}_{dec}$ of the $2^{-3}$ scale decoder.
Finally, another convolutional layer is utilized to predict the wavelet subbands $\{\mathbf{x}_{i_1, i_2, i_3}^{(-3)}\}$ in a residual learning manner, i.e., $\hat{\mathbf{x}}_{i_1, i_2, i_3}^{(-3)} = \left(\mbox{Conv}^2(\mathbf{f}^{-3}_{dec})\right)_{i_1, i_2, i_3} + \mathbf{y}_{i_1, i_2, i_3}^{(-3)}$.

In the second stage, we take $\{ \mathbf{y}_{i_1, i_2}^{(-2)} \}$ as the input to the $2^{-2}$ scale of MWRN, and $\{ \mathbf{y}_{i_1, i_2, i_3}^{(-3)} \}$ as the input to the $2^{-3}$ scale.
The scale-specific loss $\mathcal{L}^s_{-2} + \mathcal{L}^s_{-3}$ is used to train the $2^{-2}$ scale and finetune the $2^{-3}$ scale of MWRN.
As shown in Fig.~\ref{fig:step2}, we utilize a $3 \times 3$ convolutional layer on $\{ \mathbf{y}_{i_1, i_2}^{(-2)} \}$ to generate a feature map $\mathbf{F}_{\mathbf{y}}^{-2}$ of 256 channels.
Analogous to the first training stage, we assign zeros to the first level feature representation of MWRN encoder, i.e., Conv(DWT($\mathbf{f}^{-1}_{{\mathbf{y}}}$)), in the second stage.
The concatenation of $\mathbf{F}_{\mathbf{y}}^{-2}$ and Conv(DWT($\mathbf{f}^{-1}_{{\mathbf{y}}}$)) then goes through one convolutional layer and four residual blocks to generate the encoder representation $\mathbf{f}^{-2}_{{\mathbf{y}}}$ at the $2^{-2}$ scale.
Another convolutional layer is further deployed on DWT($\mathbf{f}^{-2}_{{\mathbf{y}}}$), and the output Conv(DWT($\mathbf{f}^{-2}_{{\mathbf{y}}}$)) is concatenated with $\mathbf{F}_{\mathbf{y}}^{-3}$ and is input to the $2^{-3}$ scale of MWRN.
Moreover, a convolutional layer is also applied to the decoder representation $\mathbf{f}^{-3}_{dec}$ at the $2^{-3}$ scale, where the IDWT of the output is then taken as the input to the decoder at the $2^{-2}$ scale to generate the feature representation  $\mathbf{f}^{-2}_{dec}$.
And we apply another convolutional layer on $\mathbf{f}^{-2}_{dec}$  to predict the wavelet subbands $\{\mathbf{x}_{i_1, i_2}^{(-2)}\}$.

In the third stage, we take $\{ \mathbf{y}_{i_1}^{(-1)} \}$ as the input to the $2^{-1}$ scale of MWRN, and use the loss $\mathcal{L}$ to train the $2^{-1}$ scale and finetune the $2^{-2}$ and $2^{-3}$ scales of MWRN.
As shown in Fig.~\ref{fig:step3}, the encoder feature map $\mathbf{f}_{\mathbf{y}}^{-1}$ first passes the wavelet transform and a convolutional layer is applied, which is then concatenated with $\mathbf{F}_{\mathbf{y}}^{-2}$ and input to the $2^{-2}$ scale of MWRN.
On the other hand, a convolutional layer is also applied to the decoder representation $\mathbf{f}^{-2}_{dec}$ at the $2^{-2}$ scale, where the IDWT of the output is then taken as the input to the decoder at the $2^{-1}$ scale to generate the feature representation  $\mathbf{f}^{-1}_{dec}$.
And a convolutional layer is deployed on $\mathbf{f}^{-1}_{dec}$  to predict the denoising result $\hat{\mathbf{x}}$.
At the end of training, we merge the parameters of each batch normalization into the adjacent convolution filters, and adopt a small learning rate to further finetune PT-MWRN.

Experiments are conducted on grayscale, color, and real-world noisy images to evaluate the proposed PT-MWRN.
In this section, we first describe the training and testing sets used in the experiments, and provide the implementation details of PT-MWRN.
Then, we compare our PT-MWRN with the state-of-the-art denoising methods respectively for grayscale, color and real-world image denoising.
All the source code and pre-trained models will be publicly available at \url{https://github.com/happycaoyue/PT-MWRN}.

\begin{table}[!t]\footnotesize
\centering
\caption{The setting of learning rates for our PT-MWRN. We provide the learning rates along with the epochs in each stage, and `-' denotes that the network training is early stopped in this stage. }
\label{table}
\setlength{\tabcolsep}{3pt}
\scalebox{1}{
\begin{tabular}{p{32pt}<{\centering}p{46pt}<{\centering}p{46pt}<{\centering}p{46pt}<{\centering}p{46pt}<{\centering}}
\toprule[1pt]
 &
Epoch 1$\thicksim$10&
Epoch 11$\thicksim$20&
Epoch 21$\thicksim$30&
Epoch 31$\thicksim$40\\
\bottomrule[1pt]
Stage 1 & $10^{-3}$ & $10^{-6}$ & - & -
\\

Stage 2 & $10^{-3}$ & $10^{-4}$ & $10^{-6}$ & - \\
Stage 3 & $10^{-3}$ & $10^{-4}$ & $10^{-4}$ & $10^{-6}$\\
\bottomrule[1pt]
\end{tabular}
}
\label{tab:lr}
\end{table}

\section{Experiments}
\label{sec:Experiments}

\begin{table}[!t]
\centering
\caption{Ablation studies on PT-MWRN to illustrate the effect of (i) residual blocks (RB), (ii) scale-specific input(SI), (iii) scale-specific loss (SL) and (iv) progressive training (PT).}
\label{table}
\scalebox{0.75}{
\begin{tabular}{p{60pt}<{\centering}p{20pt}<{\centering}p{20pt}<{\centering}p{20pt}<{\centering}p{20pt}<{\centering}p{30pt}<{\centering}p{35pt}<{\centering}p{35pt}<{\centering}}
\toprule[1pt]
Methods&
\#RB&
SI&
SL&
PT&
Set12&
BSD68&
Urban100\\
\bottomrule[1pt]

MWCNN&
0&
&
&
&
30.79&
29.41&
30.66\\

MWRN(2RB)&
2&
&
&
&
30.84&
29.43&
30.78\\

MWRN(4RB)&
4&
&
&
&
30.86&
29.45&
30.82\\

MWRN(6RB)&
6&
&
&
&
30.87&
29.45&
30.84\\

MWRN(SL)&
4&
&
\checkmark&
&
30.88&
29.46&
30.86\\

MWRN(SI)&
4&
\checkmark&
&
&
30.87&
29.46&
30.85\\

MWRN(SI+SL)&
4&
\checkmark&
\checkmark&
&
30.91&
29.48&
30.92\\

PT-MWRN&
4&
\checkmark&
\checkmark&
\checkmark&
30.96&
29.51&
31.00\\

\bottomrule[1pt]

\end{tabular}
}\label{tab:ab}
\end{table}


\subsection{Experimental Setting}

In this subsection, we introduce the training and testing sets used in our experiments, and describe the implementation details for training our PT-MWRN models.
Following~\cite{liu2018multi}, we use the same training set constructed by using images from three datasets, i.e., Berkeley Segmentation Dataset (BSD) \cite{martin2001database}, DIV2K~\cite{agustsson2017ntire} and Waterloo Exploration Database (WED) \cite{ma2017waterloo}, for image denoising.
Specifically, we collect $200$ images from BSD, $800$ images from DIV2K, and $4,744$ images from WED, to constitute our training set.

The grayscale and color images in the training set are respectively used to train the grayscale and color image denoising models.
For evaluating grayscale image denoising models, we adopt three testing
datasets including Set12 \cite{Zhang2016Beyond}, BSD68~\cite{martin2001database}, and Urban100~\cite{huang2015single}.
For color Gaussian noisy image denoising, we also use three datasets, i.e., CBSD68~\cite{martin2001database}, Kodak24~\cite{franzen1999kodak} and McMaster~\cite{zhang2011color}.
It should be noted that all these datasets have been widely adopted for the evaluation of Gaussian denoising methods and all the test images are not included in the training set.

\begin{table*}[!htbp]\scriptsize
\centering
\caption{
	{The average PSNR(dB)/SSIM values of gray image denoising for our PT-MWCNN and the competing methods with noise levels $\sigma$ $=$ $15$, $25$ and $50$ on the datasets
Set12, BSD68 and Urban100. {\color{red}Red}, {\color{blue}{blue}} and {\textcolor[rgb]{0.03,0.69,0.94}{cyan}} {\color{black}are utilized to indicate top {\color{red}1\textsuperscript{st}}, {\color{blue}2\textsuperscript{nd}} and  {\textcolor[rgb]{0.03,0.69,0.94}{3\textsuperscript{rd}}} rank, respectively}.}
}
\resizebox{\textwidth}{!}{
\begin{tabular}{p{20pt}<{\centering}p{6pt}<{\centering}p{42pt}<{\centering}p{42pt}<{\centering}p{44pt}<{\centering}p{44pt}<{\centering}p{42pt}<{\centering}p{46pt}<{\centering}p{46pt}<{\centering}p{44pt}<{\centering}p{42pt}<{\centering}p{44pt}<{\centering}}
\toprule[1pt]
Dataset& $\sigma$&BM3D \cite{dabov2008image} &TNRD \cite{chen2017trainable} &DnCNN \cite{Zhang2016Beyond} &IRCNN \cite{zhang2017learning} &RED \cite{mao2016image}  &MemNet \cite{tai2017memnet}&N3Net \cite{Ploetz:2018:NNN}&NLRN \cite{liu2018non} &MWCNN \cite{liu2018multi} &PT-MWRN \\
\bottomrule[1pt]
 \multirow{3}{*}{Set12}
 &15   & 32.37 / 0.8952 & 32.50 / 0.8962 & 32.86 / 0.9027 & 32.77 / 0.9008 & - & -  & -& \textcolor[rgb]{0,0,1}{33.16} / \textcolor[rgb]{0.03,0.69,0.94}{0.9070}& \textcolor[rgb]{0.03,0.69,0.94}{33.15} / \textcolor[rgb]{0,0,1}{0.9088} & \textcolor[rgb]{1,0,0}{33.30} / \textcolor[rgb]{1,0,0}{0.9110} \\
 &25   & 29.97 / 0.8505 & 30.05 / 0.8515 & 30.44 / 0.8618 & 30.38 / 0.8601 & - & - & 30.55 / 0.8668 & \textcolor[rgb]{0,0,1}{30.80} / \textcolor[rgb]{0.03,0.69,0.94}{0.8689}  & \textcolor[rgb]{0.03,0.69,0.94}{30.79} / \textcolor[rgb]{0,0,1}{0.8711} & \textcolor[rgb]{1,0,0}{30.96} / \textcolor[rgb]{1,0,0}{0.8737}\\
 &50   & 26.72 / 0.7676 & 26.82 / 0.7677 & 27.18 / 0.7827 & 27.14 / 0.7804 & 27.34 / 0.7897 & 27.38 / 0.7931 & 27.43 / 0.7931 & \textcolor[rgb]{0.03,0.69,0.94}{27.64} / \textcolor[rgb]{0.03,0.69,0.94}{0.7980} & \textcolor[rgb]{0,0,1}{27.74} / \textcolor[rgb]{0,0,1}{0.8056} & \textcolor[rgb]{1,0,0}{27.90} / \textcolor[rgb]{1,0,0}{0.8091}\\
  \hline
   \multirow{3}{*}{BSD68}
 &15   & 31.08 / 0.8722 & 31.42 / 0.8822 & 31.73 / 0.8906 & 31.63 / 0.8881 & - & - & - & \textcolor[rgb]{0,0,1}{31.88} / \textcolor[rgb]{0.03,0.69,0.94}{0.8932} & \textcolor[rgb]{0.03,0.69,0.94}{31.86} / \textcolor[rgb]{0,0,1}{0.8947} & \textcolor[rgb]{1,0,0}{31.95} / \textcolor[rgb]{1,0,0}{0.8962} \\
 &25   & 28.57 / 0.8017 & 28.92 / 0.8148 & 29.23 / 0.8278 & 29.15 / 0.8249 & - & - & 29.30 / 0.8377 & \textcolor[rgb]{0,0,1}{29.41} / \textcolor[rgb]{0.03,0.69,0.94}{0.8331} & \textcolor[rgb]{0,0,1}{29.41} / \textcolor[rgb]{0,0,1}{0.8360} & \textcolor[rgb]{1,0,0}{29.51} / \textcolor[rgb]{1,0,0}{0.8382}\\
 &50   & 25.62 / 0.6869 & 25.97 / 0.7021 & 26.23 / 0.7189 & 26.19 / 0.7171 & 26.35 / 0.7245 & 26.35 / 0.7294 & 26.39 / 0.7321 & \textcolor[rgb]{0.03,0.69,0.94}{26.47} / \textcolor[rgb]{0.03,0.69,0.94}{0.7298} & \textcolor[rgb]{0,0,1}{26.53} / \textcolor[rgb]{0,0,1}{0.7366} & \textcolor[rgb]{1,0,0}{26.62} / \textcolor[rgb]{1,0,0}{0.7399} \\
 \hline
 \multirow{3}{*}{Urban100}
 &15   & 32.34 / 0.9220 & 31.98 / 0.9187 & 32.67 / 0.9250 & 32.49 / 0.9244 & - & - & - & \textcolor[rgb]{0,0,1}{33.45} / \textcolor[rgb]{0.03,0.69,0.94}{0.9354}  & \textcolor[rgb]{0.03,0.69,0.94}{33.17} / \textcolor[rgb]{0,0,1}{0.9357} & \textcolor[rgb]{1,0,0}{33.47} / \textcolor[rgb]{1,0,0}{0.9381}\\
 &25   & 29.70 / 0.8777 & 29.29 / 0.8731 & 29.97 / 0.8792 & 29.82 / 0.8839 & - & - & 30.19 / 0.8926  & \textcolor[rgb]{0,0,1}{30.94} / \textcolor[rgb]{0.03,0.69,0.94}{0.9018}  & \textcolor[rgb]{0.03,0.69,0.94}{30.66} / \textcolor[rgb]{0,0,1}{0.9026} & \textcolor[rgb]{1,0,0}{31.00} / \textcolor[rgb]{1,0,0}{0.9033}\\
 &50   & 25.94 / 0.7791 & 25.71 / 0.7756 & 26.28 / 0.7869 & 26.14 / 0.7927 & 26.48 / 0.7991 & 26.64 / 0.8024 & 26.82 / 0.8148 & \textcolor[rgb]{0,0,1}{27.49} / \textcolor[rgb]{0.03,0.69,0.94}{0.8279} & \textcolor[rgb]{0.03,0.69,0.94}{27.42} / \textcolor[rgb]{0,0,1}{0.8371} & \textcolor[rgb]{1,0,0}{27.75} / \textcolor[rgb]{1,0,0}{0.8397} \\
\bottomrule[1pt]

\end{tabular}
}

\vspace{-0.06in}
\label{tab:gray}
\end{table*}

\begin{figure*}[!htbp]
\scriptsize{
\begin{center}
\vspace{0ex}
\subfigure{
\begin{minipage}[c]{0.19\textwidth}
\centering
  \includegraphics[width=1.03\linewidth]{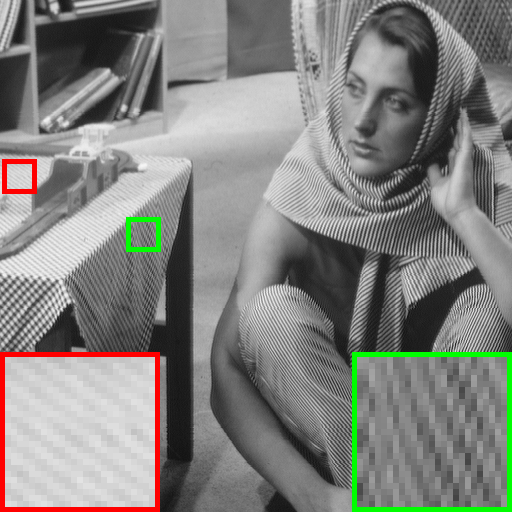}
  \centerline{\scriptsize Ground-truth}
  \label{fig:Noisy1}
\end{minipage}%
}
\subfigure{
\begin{minipage}[c]{0.19\textwidth}
\centering
  \includegraphics[width=1.03\linewidth]{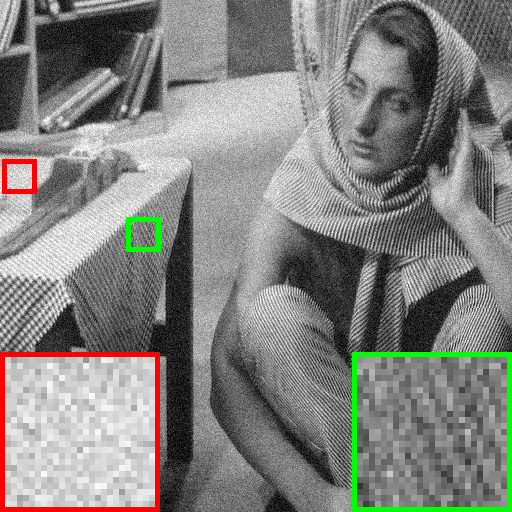}
  \centerline{\scriptsize Noisy image}
  \label{fig:Case2}
\end{minipage}%
}
\subfigure{
\begin{minipage}[c]{0.19\textwidth}
\centering
  \includegraphics[width=1.03\linewidth]{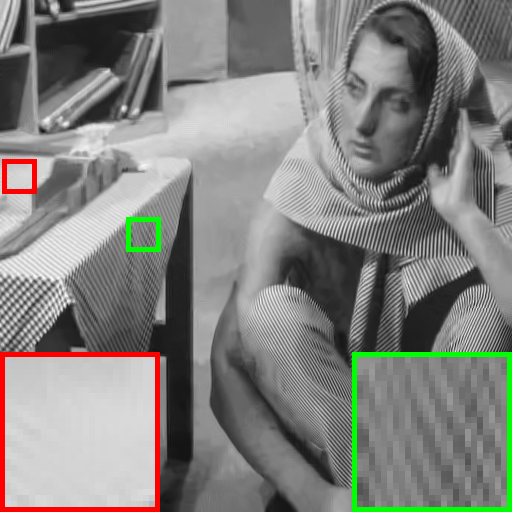}
  \centerline{\scriptsize BM3D \cite{dabov2008image}}
  \label{fig:Case2}
\end{minipage}%
}
\subfigure{
\begin{minipage}[c]{0.19\textwidth}
\centering
  \includegraphics[width=1.03\linewidth]{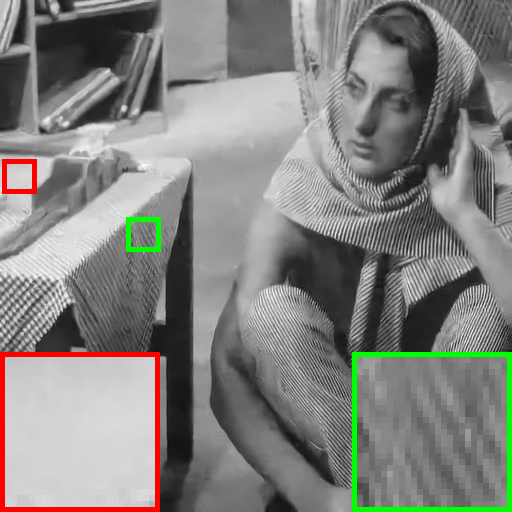}
  \centerline{\scriptsize TNRD \cite{chen2017trainable}}
  \label{fig:Case2}
\end{minipage}%
}
\subfigure{
\begin{minipage}[c]{0.19\textwidth}
\centering
  \includegraphics[width=1.03\linewidth]{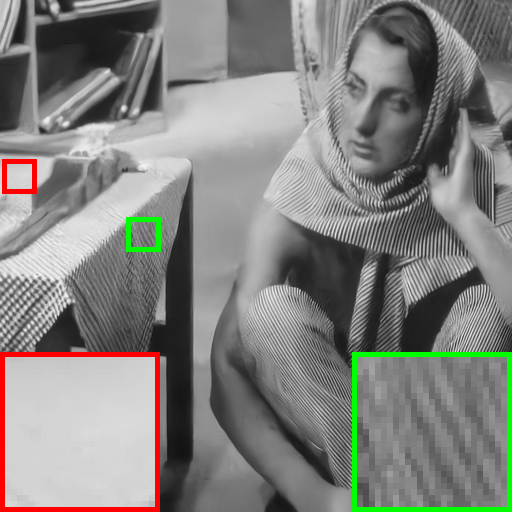}
  \centerline{\scriptsize DnCNN \cite{Zhang2016Beyond}}
  \label{fig:Case1}
\end{minipage}%
}
\vspace{-5ex}

\subfigure{
\begin{minipage}[c]{0.19\textwidth}
\centering
  \includegraphics[width=1.03\linewidth]{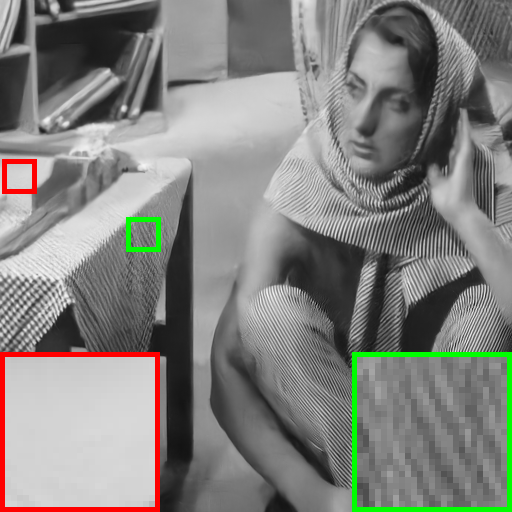}
  \centerline{\scriptsize IRCNN \cite{zhang2017learning}}
  \label{fig:Case2}
\end{minipage}%
}
\subfigure{
\begin{minipage}[c]{0.19\textwidth}
\centering
  \includegraphics[width=1.03\linewidth]{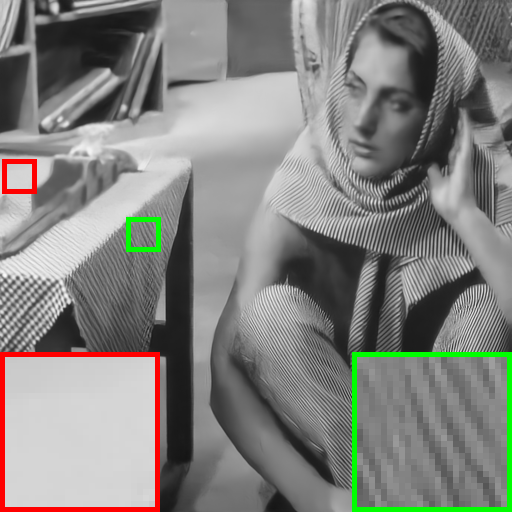}
  \centerline{\scriptsize N3Net \cite{Ploetz:2018:NNN}}
  \label{fig:comp_syn_cbsr}
\end{minipage}%
}
\subfigure{
\begin{minipage}[c]{0.19\textwidth}
\centering
  \includegraphics[width=1.03\linewidth]{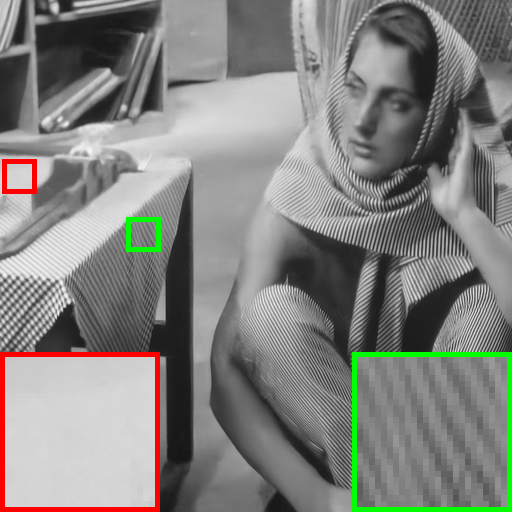}
  \centerline{\scriptsize NLRN \cite{liu2018non}}
  \label{fig:comp_syn_cbsr}
\end{minipage}%
}
\subfigure{
\begin{minipage}[c]{0.19\textwidth}
\centering
  \includegraphics[width=1.03\linewidth]{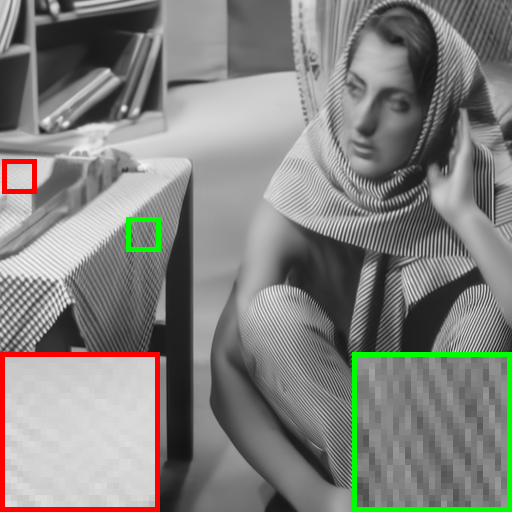}
  \centerline{\scriptsize MWCNN \cite{liu2018multi}}
  \label{fig:Case2}
\end{minipage}%
}
\subfigure{
\begin{minipage}[c]{0.19\textwidth}
\centering
  \includegraphics[width=1.03\linewidth]{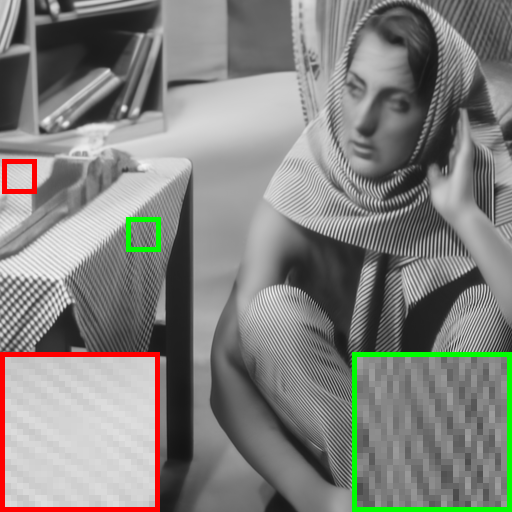}
  \centerline{\scriptsize PT-MWRN}
  \label{fig:Case2}
\end{minipage}%
}
\vspace{-3ex}
\caption{\small Gaussian denoising results of the grayscale image ``09" from Set12 with noise level 25.
}\label{fig:f2}
\vspace{-1ex}
\end{center}}
\end{figure*}


\begin{figure*}[!htbp]
\scriptsize{
\begin{center}
\vspace{-0ex}
\subfigure{
\begin{minipage}[c]{0.15\textwidth}
\centering
  \includegraphics[width=1.05\linewidth]{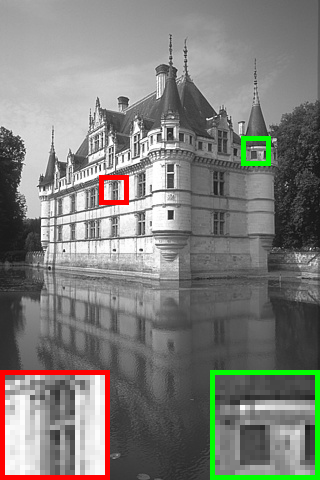}
  \centerline{\scriptsize Ground-truth}
  \label{fig:comp_syn_lrbic}
\end{minipage}%
}
\subfigure{
\begin{minipage}[c]{0.15\textwidth}
\centering
  \includegraphics[width=1.05\linewidth]{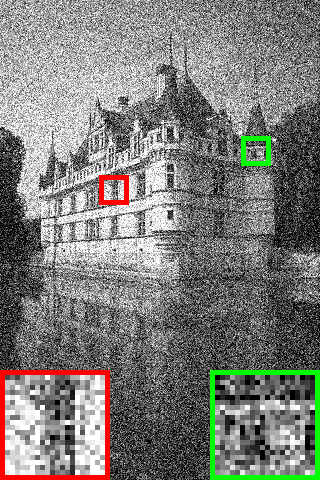}
  \centerline{\scriptsize Noisy image}
  \label{fig:comp_syn_dbpn}
\end{minipage}%
}
\subfigure{
\begin{minipage}[c]{0.15\textwidth}
\centering
  \includegraphics[width=1.05\linewidth]{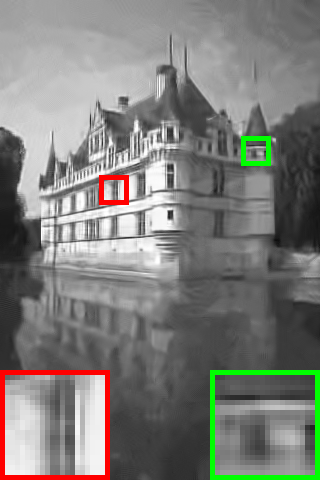}
  \centerline{\scriptsize BM3D \cite{dabov2008image}}
  \label{fig:comp_syn_edsr}
\end{minipage}%
}
\subfigure{
\begin{minipage}[c]{0.15\textwidth}
\centering
  \includegraphics[width=1.05\linewidth]{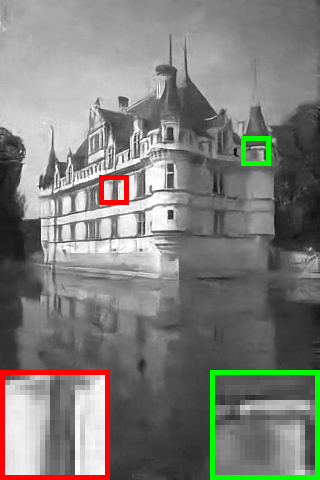}
  \centerline{\scriptsize TNRD \cite{chen2017trainable}}
  \label{fig:comp_syn_rcan}
\end{minipage}%
}
\subfigure{
\begin{minipage}[c]{0.15\textwidth}
\centering
  \includegraphics[width=1.05\linewidth]{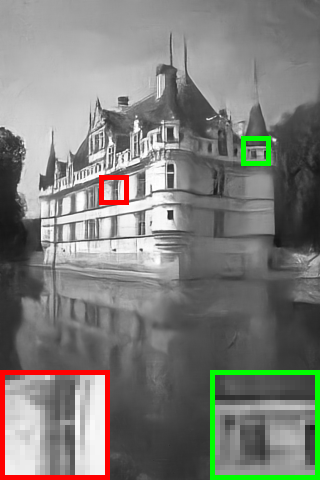}
  \centerline{\scriptsize DnCNN \cite{Zhang2016Beyond}}
  \label{fig:comp_syn_srresnet+}
\end{minipage}%
}
\subfigure{
\begin{minipage}[c]{0.15\textwidth}
\centering
  \includegraphics[width=1.05\linewidth]{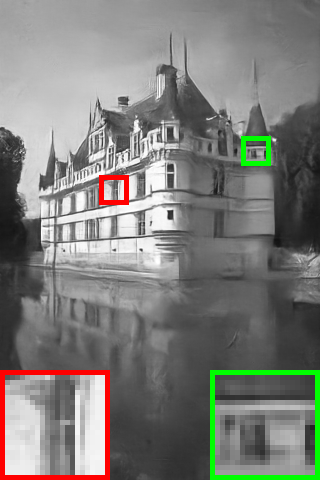}
  \centerline{\scriptsize IRCNN \cite{zhang2017learning}}
  \label{fig:comp_syn_rcan+}
\end{minipage}%
}
\vspace{-5ex}

\subfigure{
\begin{minipage}[c]{0.15\textwidth}
\centering
  \includegraphics[width=1.05\linewidth]{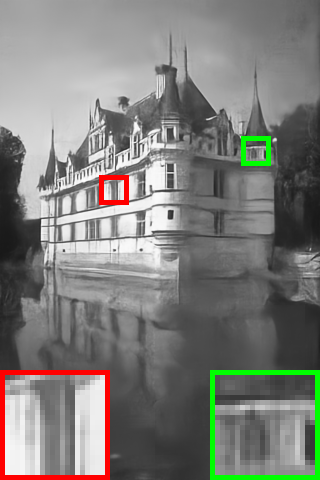}
  \centerline{\scriptsize RED \cite{mao2016image}}
  \label{fig:comp_syn_srmd}
\end{minipage}%
}
\subfigure{
\begin{minipage}[c]{0.15\textwidth}
\centering
  \includegraphics[width=1.05\linewidth]{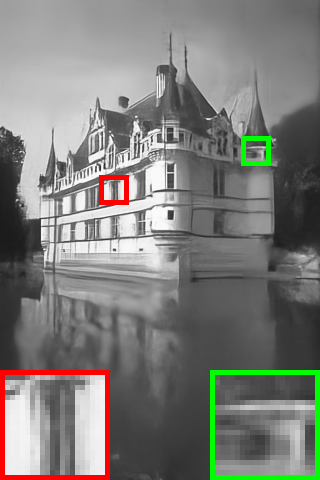}
  \centerline{\scriptsize MemNet \cite{tai2017memnet} }
  \label{fig:comp_syn_dbrcan}
\end{minipage}%
}
\subfigure{
\begin{minipage}[c]{0.15\textwidth}
\centering
  \includegraphics[width=1.05\linewidth]{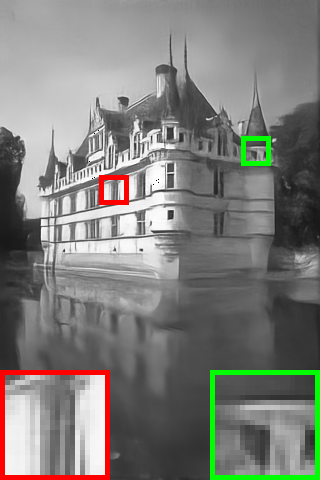}
  \centerline{\scriptsize N3Net \cite{Ploetz:2018:NNN}}
  \label{fig:comp_syn_cbsr}
\end{minipage}%
}
\subfigure{
\begin{minipage}[c]{0.15\textwidth}
\centering
  \includegraphics[width=1.05\linewidth]{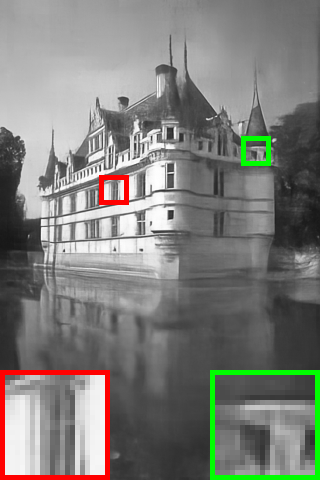}
  \centerline{\scriptsize NLRN \cite{liu2018non}}
  \label{fig:comp_syn_cbsr}
\end{minipage}%
}
\subfigure{
\begin{minipage}[c]{0.15\textwidth}
\centering
  \includegraphics[width=1.05\linewidth]{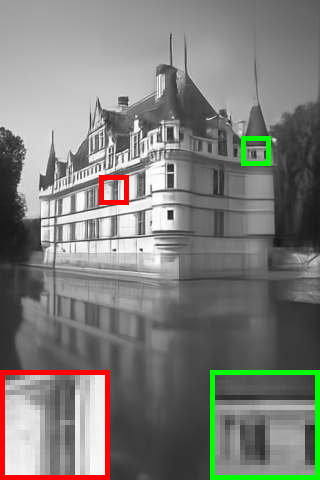}
  \centerline{\scriptsize MWCNN \cite{liu2018multi}}
  \label{fig:comp_syn_cbsr}
\end{minipage}%
}
\subfigure{
\begin{minipage}[c]{0.15\textwidth}
\centering
  \includegraphics[width=1.05\linewidth]{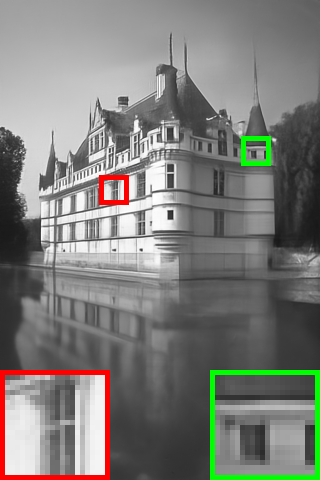}
  \centerline{\scriptsize PT-MWRN}
  \label{fig:comp_syn_gt}
\end{minipage}%
}
\vspace{-3ex}
\caption{\small
{Gaussian denoising results of the grayscale image ``Test003" from BSD68 with noise level 50.}}
\label{fig:f3}
\vspace{-1ex}
\end{center}}
\end{figure*}
%

\begin{table*}[!t]\footnotesize
\centering
\caption{
	{ The average PSNR(dB)/SSIM values of color image denoising for our PT-MWRN and the competing methods with noise levels $\sigma$ $=$ $15$, $25$ and $50$ on the
CBSD68, Kodak24 and McMaster datasets. {\color{red}Red}, {\color{blue}{blue}} and {\textcolor[rgb]{0.03,0.69,0.94}{cyan}} {\color{black}are utilized to indicate top {\color{red}1\textsuperscript{st}}, {\color{blue}2\textsuperscript{nd}} and  {\textcolor[rgb]{0.03,0.69,0.94}{3\textsuperscript{rd}}} rank, respectively}}
}

\scalebox{1}{\begin{tabular}{p{27pt}<{\centering}p{8pt}<{\centering}p{50pt}<{\centering}p{50pt}<{\centering}p{50pt}<{\centering}p{48pt}<{\centering}p{48pt}<{\centering}p{51pt}<{\centering}p{55pt}<{\centering}}
\toprule[1pt]
Dataset& $\sigma$&CBM3D \cite{dabov2008image} &IRCNN \cite{zhang2017learning} &CDnCNN \cite{Zhang2016Beyond} &FFDNet \cite{zhang2018ffdnet} & DHDN \cite{park2019densely} &CMWCNN \cite{liu2018multi}  &PT-MWRN \\
\bottomrule[1pt]
 \multirow{3}{*}{CBSD68}
 &15   & 33.52 / 0.9215 & 33.86 / 0.9283 & \textcolor[rgb]{0.03,0.69,0.94}{33.89} / \textcolor[rgb]{0.03,0.69,0.94}{0.9289}  & 33.84 / \textcolor[rgb]{0.03,0.69,0.94}{0.9289} & - &  \textcolor[rgb]{0,0,1}{34.08} /  \textcolor[rgb]{0,0,1}{0.9316} & \textcolor[rgb]{1,0,0}{34.22} / \textcolor[rgb]{1,0,0}{0.9337} \\
 &25   & 30.71 / 0.8672 & 31.17 / 0.8821 & \textcolor[rgb]{0.03,0.69,0.94}{31.23} / 0.8826  & 31.21 / \textcolor[rgb]{0.03,0.69,0.94}{0.8831} & - &  \textcolor[rgb]{0,0,1}{31.40} /  \textcolor[rgb]{0,0,1}{0.8873}& \textcolor[rgb]{1,0,0}{31.58} / \textcolor[rgb]{1,0,0}{0.8906}  \\
 &50   & 27.38 / 0.7626 & 27.86 / 0.7889 & 27.92 / 0.7885  & 27.96 / 0.7911 & \textcolor[rgb]{0.03,0.69,0.94}{28.05}/\textcolor[rgb]{0.03,0.69,0.94}{0.7951} &  \textcolor[rgb]{0,0,1}{28.26} /  \textcolor[rgb]{0,0,1}{0.8031} & \textcolor[rgb]{1,0,0}{28.40} / \textcolor[rgb]{1,0,0}{0.8062}  \\
  \hline
   \multirow{3}{*}{Kodak24}
 &15   & 34.28 / 0.9147 & 34.55 / 0.9198 & 34.47 / 0.9198 & \textcolor[rgb]{0.03,0.69,0.94}{34.61} / \textcolor[rgb]{0.03,0.69,0.94}{0.9217} & - &  \textcolor[rgb]{0,0,1}{34.88} /  \textcolor[rgb]{0,0,1}{0.9250} & \textcolor[rgb]{1,0,0}{35.08} / \textcolor[rgb]{1,0,0}{0.9277} \\
 &25   & 31.68 / 0.8670 & 32.02 / 0.8767 & 32.03 / 0.8764& \textcolor[rgb]{0.03,0.69,0.94}{32.13} / \textcolor[rgb]{0.03,0.69,0.94}{0.8789} & - &  \textcolor[rgb]{0,0,1}{32.38} /  \textcolor[rgb]{0,0,1}{0.8845} & \textcolor[rgb]{1,0,0}{32.62}  / \textcolor[rgb]{1,0,0}{0.8990} \\
 &50   & 28.46 / 0.7760 & 28.80 / 0.7933 & 28.83  / 0.7915 & 28.97 / 0.7970 &  \textcolor[rgb]{0,0,1}{29.40}/ \textcolor[rgb]{0,0,1}{0.8123} & \textcolor[rgb]{0.03,0.69,0.94}{29.37} / \textcolor[rgb]{0.03,0.69,0.94}{0.8115} & \textcolor[rgb]{1,0,0}{29.58} / \textcolor[rgb]{1,0,0}{0.8162} \\ \hline
 \multirow{3}{*}{McMaster}
 &15   & 34.06 / 0.9114 & 34.57 / 0.9196 & 33.44 / 0.9035 & \textcolor[rgb]{0.03,0.69,0.94}{34.70} / \textcolor[rgb]{0.03,0.69,0.94}{0.9220} & - &  \textcolor[rgb]{0,0,1}{34.92} /  \textcolor[rgb]{0,0,1}{0.9261}& \textcolor[rgb]{1,0,0}{35.20} / \textcolor[rgb]{1,0,0}{0.9297} \\
 &25   & 31.66 / 0.8699 & 32.18 / 0.8819 & 31.52 / 0.8694 & \textcolor[rgb]{0.03,0.69,0.94}{32.37} / \textcolor[rgb]{0.03,0.69,0.94}{0.8874} & - &  \textcolor[rgb]{0,0,1}{32.55} /  \textcolor[rgb]{0,0,1}{0.8924} & \textcolor[rgb]{1,0,0}{32.89} / \textcolor[rgb]{1,0,0}{0.8979} \\
 &50   & 28.51 / 0.7911 & 28.92 / 0.8068 & 28.62 / 0.7989 & 29.19 / 0.8192 & \textcolor[rgb]{0.03,0.69,0.94}{29.59} / \textcolor[rgb]{0,0,1}{0.8248} &  \textcolor[rgb]{0,0,1}{29.61} / \textcolor[rgb]{0.03,0.69,0.94}{0.8323}& \textcolor[rgb]{1,0,0}{29.87} / \textcolor[rgb]{1,0,0}{0.8375} \\
\bottomrule[1pt]

\end{tabular}
}

\vspace{-0.06in}
\label{tab:color}
\end{table*}

We randomly crop $24 \times 6,000$ patches with the size of $240 \times 240$ from the training images.
To train PT-MWRN for Gaussian denoising with known noise level, we consider three noise levels $\sigma$ $=$ $15$, $25$ and $50$.
We use the ADAM \cite{kingma2015adam} optimizer with $\beta_1 = 0.9$, $\beta_2 = 0.999$ and $\epsilon = 1e^{-8}$ to minimize the loss function.
As to the other hyper-parameters of ADAM, the default setting is adopted.
Generally, the training of PT-MWRN involves three stages, and at most $40$ epochs are adopted in each stage.
The learning rate is set as Table~\ref{tab:lr} and we adopt the mini-batch size $24$ during training.
Rotation or/and flip based data augmentation is used during training.
We train and test our network using the MatConvNet package \cite{vedaldi2015matconvnet}.
All the experiments are carried out in the Windows 10 and Matlab (R2017b) environment running on a PC with Xeon E3-1230 v5 CPU, Nvidia GeForce GTX 1080 Ti GPU and 32 GB RAM.
And the Nvidia CUDA 8.0 and cuDNN-v5.1 are utilized to accelerate the GPU computation.
It takes about two days to train a PT-MWRN model.

\subsection{Ablation Study}

Ablation study is conducted on the grayscale images to evaluate our PT-MWRN.
Generally, our PT-MWRN improves MWCNN from four aspects:
(i) replacing the convolution layers with residual blocks (RB),
(ii) feeding the wavelet subbands of noisy image into each scale of the
encoder (i.e., scale-specific input, SI),
(iii) incorporating scale-specific loss (i.e., SL),
and (iv) progressive training (PT).
To illustrate the effect of these components, we implement seven variants of PT-WMRN, i.e., MWRN(2RB), MWRN(4RB), MWRN(6RB), MWRN(SI), MWRN(SL), MWRN(SI+SL), and the full PT-MWRN.
The configuration of each variant is provided in Table~\ref{tab:ab}, and we also take MWCNN as a baseline method in the ablation study.

%
%

Using AWGN with $\sigma = 25$ as an example, Table~\ref{tab:ab} lists the PSNR and SSIM results of the PT-MWRN variants on the three grayscale image datasets, i.e., Set12, BSD68, and Urban100.
%
%
In contrast to MWCNN, MWRN(4RB) achieves 0.04 dB gain by PSNR on BSD68, and further increasing the number of residual blocks contributes little to performance improvement.
Thus, four residual blocks are adopted in the default PT-MWRN.
Moreover, the incorporation of scale-specific input and scale-specific loss, i.e., MWRN(SI+SL), can bring another 0.04 dB gain against MWRN(4RB).
In comparison, the improvements by MWRN(SI) and MWRN(SL) are much smaller, i.e., only 0.01 dB, clearly indicating that scale-specific input and scale-specific loss are complementary and can be combined to boost denoising performance.
Finally, progressive training can bring another 0.03 dB gain against MWRN(SI+SL).
To sum up, benefited from residual blocks, scale-specific loss and progressive training, our PT-MWRN can achieve PSNR gains of 0.17, 0.10, and 0.34 dB against MWCNN for Gaussian denoising with $\sigma = 25$ respectively on the Set12, BSD68, and Urban100 datasets.


\subsection{Experiments on Grayscale Image Denoising}

For Gaussian denoising of grayscale images, we compare our PT-MWRN with nine competing denoising methods, i.e., BM3D \cite{dabov2008image}, TNRD \cite{chen2017trainable}, DnCNN \cite{Zhang2016Beyond}, IRCNN \cite{zhang2017learning}, RED \cite{mao2016image}, MemNet \cite{tai2017memnet}, N3Net \cite{Ploetz:2018:NNN}, NLRN \cite{liu2018non} and MWCNN \cite{liu2018multi}.
Table~\ref{tab:gray} lists the average PSNR(dB)/SSIM results of different methods on Set12 \cite{Zhang2016Beyond}, BSD68\cite{martin2001database}, and Urban100 \cite{huang2015single}.
It can be seen that our PT-MWRN performs favorably against all the competing methods for all the noise levels and datasets.
We also note that our PT-MWRN surpasses MWCNN $0.1$ dB in terms of PSNR on BSD68 and near about 0.2 dB on Set12, indicating the effectiveness of incorporating deep MWRN architecture, scale-specific loss and progressive training.

Visual comparison of denoising results is further provided.
Figs.~\ref{fig:f2} and~\ref{fig:f3} show the denoising results of two grayscale images respectively from Set12 \cite{Zhang2016Beyond} ($\sigma$ $=$ $25$) and BSD68 \cite{martin2001database} ($\sigma$ $=$ $50$).
It can be seen that some small-scale textural details are blurred or smoothed out for the results by BM3D \cite{dabov2008image}, TNRD \cite{chen2017trainable}, DnCNN \cite{Zhang2016Beyond}, IRCNN \cite{zhang2017learning}, RED \cite{mao2016image}, MemNet \cite{tai2017memnet}, N3Net \cite{Ploetz:2018:NNN}.
MWCNN \cite{liu2018multi} and NLRN \cite{liu2018non} are effective in preserving fine details.
In comparison to the competing methods, our PT-MWRN is better at recovering image details and structures, and can obtain visually more pleasing result.
%

\subsection{Experiments on Color Image Denoising}
\begin{figure*}[!htbp]
\scriptsize{
\begin{center}
\vspace{0ex}
\subfigure{
\begin{minipage}[c]{0.22\textwidth}
\centering
  \includegraphics[width=1.03\linewidth]{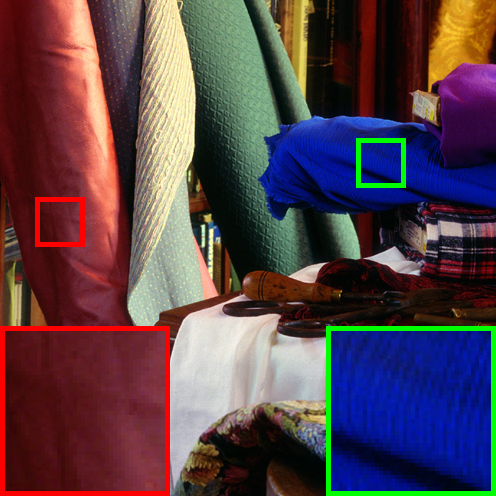}
  \centerline{\scriptsize Ground-truth}
  \label{fig:Noisy1}
\end{minipage}%
}
\subfigure{
\begin{minipage}[c]{0.22\textwidth}
\centering
  \includegraphics[width=1.03\linewidth]{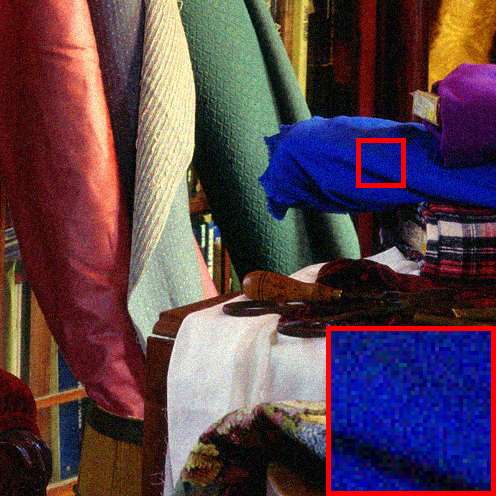}
  \centerline{\scriptsize Noisy image}
  \label{fig:Case2}
\end{minipage}%
}
\subfigure{
\begin{minipage}[c]{0.22\textwidth}
\centering
  \includegraphics[width=1.03\linewidth]{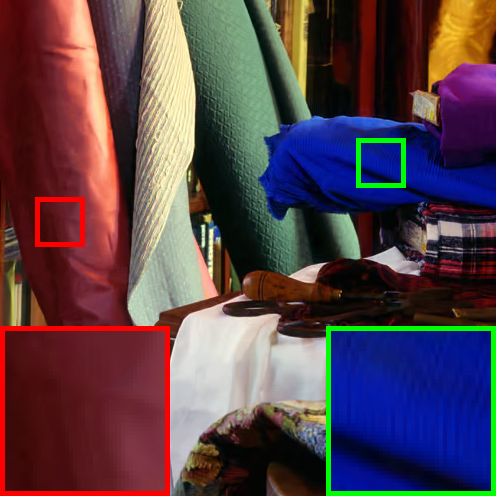}
  \centerline{\scriptsize CBM3D \cite{dabov2008image}}
  \label{fig:Case2}
\end{minipage}%
}
\subfigure{
\begin{minipage}[c]{0.22\textwidth}
\centering
  \includegraphics[width=1.03\linewidth]{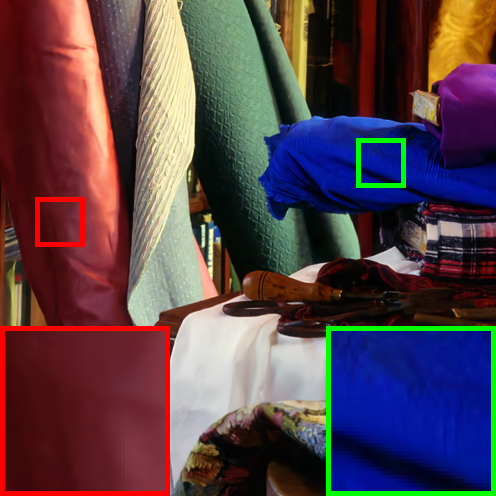}
  \centerline{\scriptsize FFDnet \cite{zhang2018ffdnet}}
  \label{fig:Case2}
\end{minipage}%
}

\vspace{-5ex}

\subfigure{
\begin{minipage}[c]{0.22\textwidth}
\centering
  \includegraphics[width=1.03\linewidth]{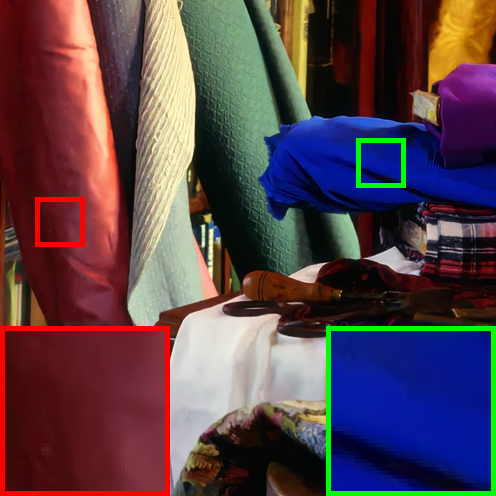}
  \centerline{\scriptsize CDnCNN \cite{Zhang2016Beyond}}
  \label{fig:Case1}
\end{minipage}%
}
\subfigure{
\begin{minipage}[c]{0.22\textwidth}
\centering
  \includegraphics[width=1.03\linewidth]{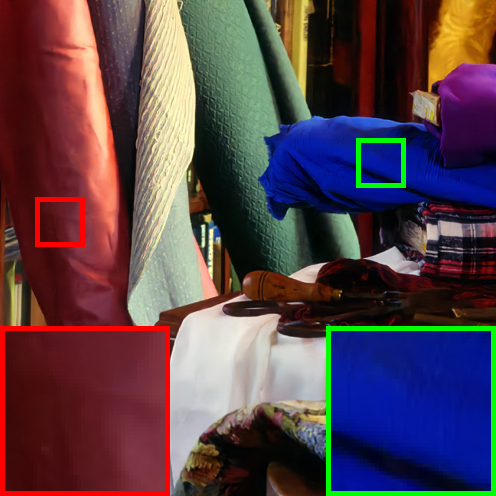}
  \centerline{\scriptsize IRCNN \cite{zhang2017learning}}
  \label{fig:Case2}
\end{minipage}%
}
\subfigure{
\begin{minipage}[c]{0.22\textwidth}
\centering
  \includegraphics[width=1.03\linewidth]{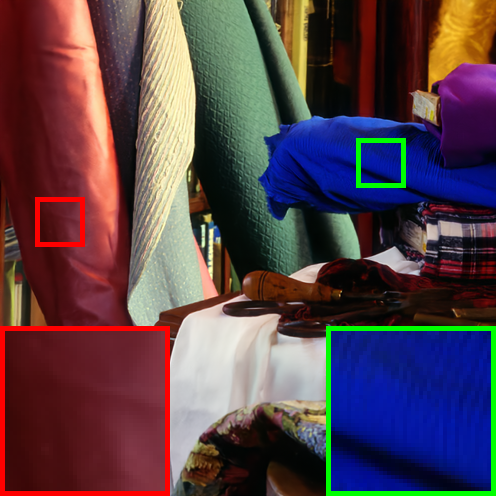}
  \centerline{\scriptsize CMWCNN \cite{liu2018multi}}
  \label{fig:Case2}
\end{minipage}%
}
\subfigure{
\begin{minipage}[c]{0.22\textwidth}
\centering
  \includegraphics[width=1.03\linewidth]{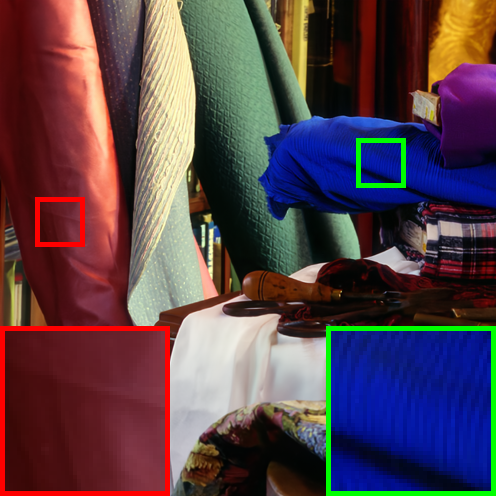}
  \centerline{\scriptsize PT-MWRN}
  \label{fig:Case2}
\end{minipage}%
}

\vspace{-3ex}

\caption{\small Denoising results of the color image ``09" from the McMaster12 dataset with noise level 15.
}\label{fig:f4}
\vspace{-4ex}
\end{center}}
\end{figure*}

\begin{figure*}[!htbp]
\scriptsize{
\begin{center}
\vspace{0ex}
\subfigure{
\begin{minipage}[c]{0.2\textwidth}
\centering
  \includegraphics[width=1.03\linewidth]{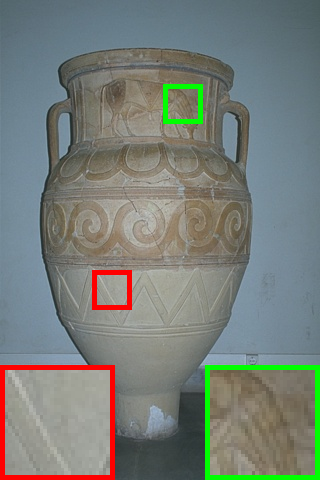}
  \centerline{\scriptsize Ground-truth}
  \label{fig:Noisy1}
\end{minipage}%
}
\subfigure{
\begin{minipage}[c]{0.2\textwidth}
\centering
  \includegraphics[width=1.03\linewidth]{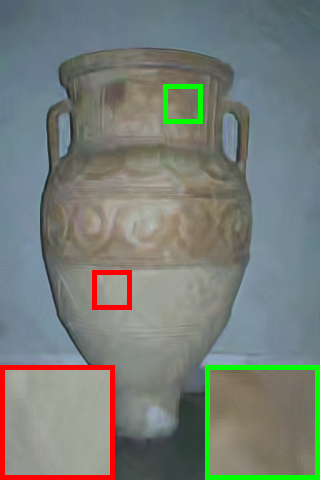}
  \centerline{\scriptsize CBM3D \cite{dabov2008image}}
  \label{fig:Case2}
\end{minipage}%
}
\subfigure{
\begin{minipage}[c]{0.2\textwidth}
\centering
  \includegraphics[width=1.03\linewidth]{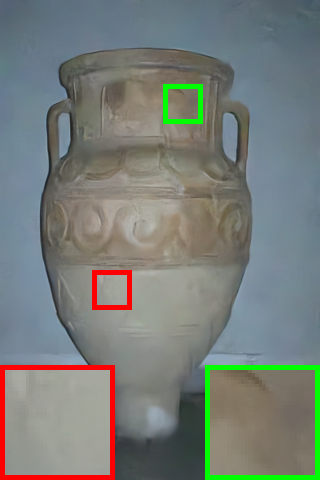}
  \centerline{\scriptsize FFDnet \cite{zhang2018ffdnet}}
  \label{fig:Case2}
\end{minipage}%
}
\subfigure{
\begin{minipage}[c]{0.2\textwidth}
\centering
  \includegraphics[width=1.03\linewidth]{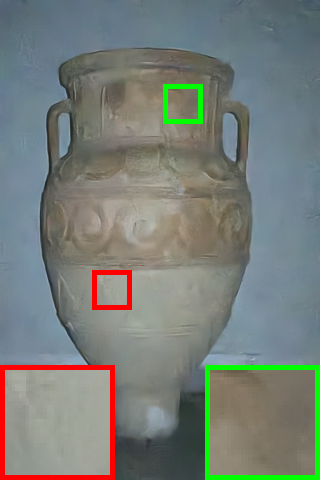}
  \centerline{\scriptsize CDnCNN \cite{Zhang2016Beyond}}
  \label{fig:Case2}
\end{minipage}%
}

\vspace{-5ex}

\subfigure{
\begin{minipage}[c]{0.2\textwidth}
\centering
  \includegraphics[width=1.03\linewidth]{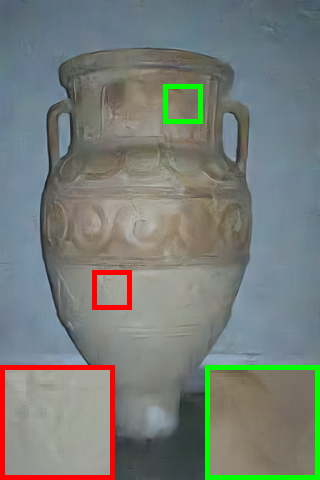}
  \centerline{\scriptsize IRCNN \cite{zhang2017learning}}
  \label{fig:Case1}
\end{minipage}%
}
\subfigure{
\begin{minipage}[c]{0.2\textwidth}
\centering
  \includegraphics[width=1.03\linewidth]{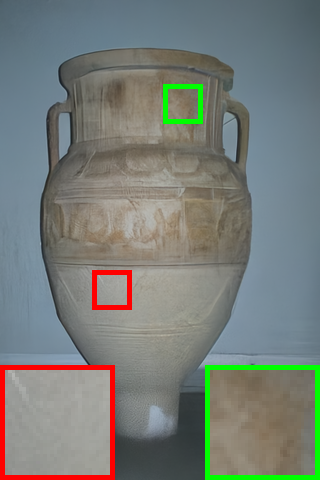}
  \centerline{\scriptsize DHDN \cite{park2019densely}}
  \label{fig:Case2}
\end{minipage}%
}
\subfigure{
\begin{minipage}[c]{0.2\textwidth}
\centering
  \includegraphics[width=1.03\linewidth]{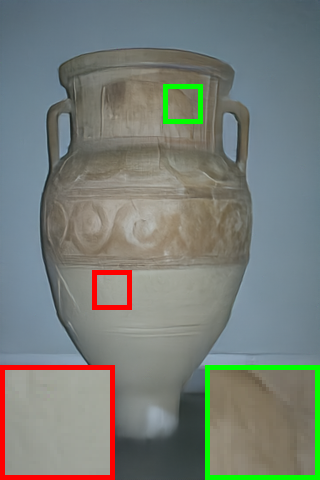}
  \centerline{\scriptsize CMWCNN \cite{liu2018multi}}
  \label{fig:Case2}
\end{minipage}%
}
\subfigure{
\begin{minipage}[c]{0.2\textwidth}
\centering
  \includegraphics[width=1.03\linewidth]{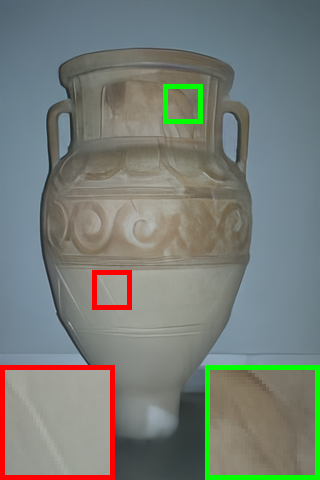}
  \centerline{\scriptsize PT-MWRN}
  \label{fig:Case2}
\end{minipage}%
}

\vspace{-3ex}

\caption{\small Denoising results of the color image ``227092" from CBSD68 with noise level 50.
}\label{fig:f5}
\vspace{-4ex}
\end{center}}
\end{figure*}

In addition to grayscale image denoising, we also train the color Gaussian image denoising model.
MWCNN~\cite{liu2018multi} is only trained on grayscale images in the original paper, and here we retrain MWCNN for color image denoising, dubbed by CMWCNN.
For color image denoising, we compare our PT-MWRN with five competing methods, i.e., CBM3D \cite{dabov2008image}, IRCNN \cite{zhang2017learning}, CDnCNN \cite{Zhang2016Beyond}, FFDNet \cite{zhang2018ffdnet}, DHDN \cite{park2019densely} and CMWCNN \cite{liu2018multi}.
Table~\ref{tab:color} lists the average PSNR(dB)/SSIM results of different methods on the CBSD68, Kodak24 and McMaster datasets.
In terms of both PSNR and SSIM, our PT-MWRN outperforms the competing methods on the three datasets.
On the Kodak24 dataset, the PSNR values of our PT-MWRN are more than 0.34dB higher than those of the second best method, i.e., CMWCNN~\cite{liu2018multi}, for AWGN with any noise levels.
The quantitative results further demonstrate the effectiveness of our PT-MWRN for color image denoising.
Figs.~\ref{fig:f4} and~\ref{fig:f5} respectively show the denoising results of two images by different methods.
And our PT-MWRN can yield cleaner images with much finer details.



\subsection{Running Time}

We test the running time of our PT-MWRN as well as the competing methods on grayscale image denoising.
Table~\ref{tab:time} gives the running time for different image sizes.
In particular, PT-MWRN is much faster than RED \cite{mao2016image} and MemNet \cite{tai2017memnet}, but is less efficient than TNRD \cite{chen2017trainable}, DnCNN \cite{Zhang2016Beyond} and MWCNN \cite{liu2018multi}.
Taking the PSNR/SSIM gains against the competing methods into account, our PT-MWRN can be regarded as a suitable tradeoff between denoising performance and computational efficiency.

%
%

\begin{table}[!t]
\centering
\caption{Running time (in seconds) of the competing methods on images of size $256 \times 256$, $512 \times 512$ and $1024 \times 1024$.}
\label{table}

\scalebox{0.55}{
\begin{tabular}{p{1.8cm}<{\centering}p{1.8cm}<{\centering}p{2cm}<{\centering}p{1.6cm}<{\centering}p{2.0cm}<{\centering}p{2.0cm}<{\centering}p{1.8cm}<{\centering}}
\toprule[1pt]
Image Size&
TNRD \cite{chen2017trainable}&
DnCNN \cite{Zhang2016Beyond}&
RED \cite{mao2016image}&
MemNet \cite{tai2017memnet} &
MWCNN \cite{liu2018multi}&
PT-MWRN\\
\bottomrule[1pt]
$256 \times 256$&
0.013&
0.0193&
1.421&
0.9121&
0.0640&
0.0996\\

$512 \times 512$&
0.037&
0.0563&
4.912&
3.804&
0.0983&
0.1457\\

$1024 \times 1024$&
0.121&
0.1801&
16.05&
14.92&
0.3646&
0.5731\\
\bottomrule[1pt]
\end{tabular}
}
\label{tab:time}
\end{table}

\begin{table}[!t]
\centering
\caption{The mean PSNR and SSIM results of state-of-the-art algorithms evaluated on the DND benchmark. {\color{red}Red}, {\color{blue}{blue}} and {\textcolor[rgb]{0.03,0.69,0.94}{cyan}} {\color{black}are utilized to indicate top {\color{red}1\textsuperscript{st}}, {\color{blue}2\textsuperscript{nd}} and  {\textcolor[rgb]{0.03,0.69,0.94}{3\textsuperscript{rd}}} rank, respectively.} }
\label{table}

\scalebox{0.64}{
\begin{tabular}{p{3.4cm}<{\centering}p{2.6cm}<{\centering}p{2cm}<{\centering}p{1.6cm}<{\centering}p{2cm}<{\centering}}
\toprule[1pt]
Method&
Blind/Non-blind&
Denoising on&
PSNR&
SSIM\\
\bottomrule[1pt]
CDnCNN-B~\cite{Zhang2016Beyond}&
blind&
sRGB&
32.43&
0.7900\\

TNRD \cite{chen2017trainable}&
Non-blind&
sRGB&
33.65&
0.8306\\

FFDNet~\cite{zhang2018ffdnet}&
Non-blind&
sRGB&
34.40&
0.8474\\

BM3D~\cite{dabov2008image}&
Non-blind&
sRGB&
34.51&
0.8507\\

KSVD~\cite{aharon2006k}&
Non-blind&
sRGB&
36.49&
0.8507\\

MCWNNM~\cite{xu2017multi}&
blind&
sRGB&
37.38&
0.9294\\

TWSC \cite{xu2018trilateral}&
blind&
sRGB&
37.94&
0.9403\\

CBDNet \cite{guo2019toward}&
blind&
sRGB&
38.06&
0.8978\\

RIDNet \cite{anwar2019real}&
blind&
sRGB&
39.26&
0.9528\\

VDN \cite{yue2019variational}&
blind&
sRGB&
39.38&
0.9518\\

PRIDNet \cite{zhao2019pyramid}&
blind&
sRGB&
39.42&
0.9528\\

CycleISP \cite{Zamir2020CycleISP}&
blind&
sRGB&
39.56&
0.9560\\

Path-Restore \cite{yu2019path}&
blind&
sRGB&
\textcolor[rgb]{0.03,0.69,0.94}{39.72}&
\textcolor[rgb]{1,0,0}{0.9591}\\

AINDNet \cite{Kim2020Aindnet}&
blind&
sRGB&
\textcolor[rgb]{0,0,1}{39.77}&
\textcolor[rgb]{0,0,1}{0.9590}\\

PT-MWRN(Ours)&
blind&
sRGB&
\textcolor[rgb]{1,0,0}{39.84}&
\textcolor[rgb]{0.03,0.69,0.94}{0.9580}\\

\bottomrule[1pt]
\end{tabular}
}
\label{tab:realDND}
\vspace{-0.145in}
\end{table}

\subsection{Blind Denoising of Real-world Noisy Photographs}
When trained with the pairs of real-world noisy and clean images, our PT-MWRN can also be utilized for the denoising of real-world photographs.
To this end, we combine the SIDD Medium Dataset \cite{SIDD_2018_CVPR} and synthetic realistic noisy images adopted in~\cite{guo2019toward} as our training data.
%
%
Concretely, we employ the noise level estimation subnetwork in CBDNet~\cite{guo2019toward} and adopt PT-MWRN as the denoising subnetwork, which are then jointly trained using the training set.
%
%
%
The DND \cite{plotz2017benchmarking} and SIDD \cite{SIDD_2018_CVPR} benchmarks are used to test our PT-MWRN on real-world noisy photographs.

\textbf{DND.} The DND benchmark \cite{plotz2017benchmarking} contains 50 testing real-world noisy photographs, and does not provide any additional training data for fine-tuning denoising networks.
Using PSNR and SSIM as performance metrics, we compare our PT-MWRN with the existing competitive algorithms including both traditional algorithms BM3D \cite{dabov2008image}, KSVD \cite{aharon2006k}, MCWNNM \cite{xu2017multi}, TWSC \cite{xu2018trilateral} and deep denoisers CDnCNN-B\cite{Zhang2016Beyond}, TNRD \cite{chen2017trainable}, FFDNet\cite{zhang2018ffdnet}, CBDNet \cite{guo2019toward}, RIDNet \cite{anwar2019real}, PRIDNet \cite{zhao2019pyramid}, VDN \cite{yue2019variational}, CycleISP \cite{Zamir2020CycleISP}, Path-Restore \cite{yu2019path} and AINDNet \cite{Kim2020Aindnet}.
%
%
Table~\ref{tab:realDND} lists the denoising performance of the competing methods.
As the clean ground-truth images are not released, all the listed results are from the official benchmark website\footnote{\url{https://noise.visinf.tu-darmstadt.de/benchmark/}}.
CDnCNN-B \cite{Zhang2016Beyond} and FFDNet\cite{zhang2018ffdnet} are trained using synthetic AWGN and thus perform poorly in handling real-world noisy photographs.
In term of PSNR, our PT-MWRN outperforms all the other approaches including both traditional and deep denoisers.
The SSIM values of Path-Restore \cite{yu2019path}, AINDNet \cite{Kim2020Aindnet} and our PT-MWRN are comparable, and are higher than those of the other approaches.

%

\begin{figure*}[!htbp]
\scriptsize{
\begin{center}
\vspace{+1ex}
\subfigure{
\begin{minipage}[c]{0.2\textwidth}
\centering
  \includegraphics[width=1.03\linewidth]{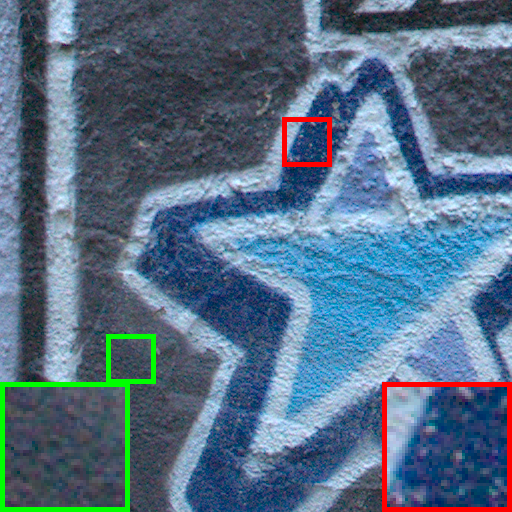}
  \centerline{\scriptsize Noisy}
  \label{fig:Noisy1}
\end{minipage}%
}
\subfigure{
\begin{minipage}[c]{0.2\textwidth}
\centering
  \includegraphics[width=1.03\linewidth]{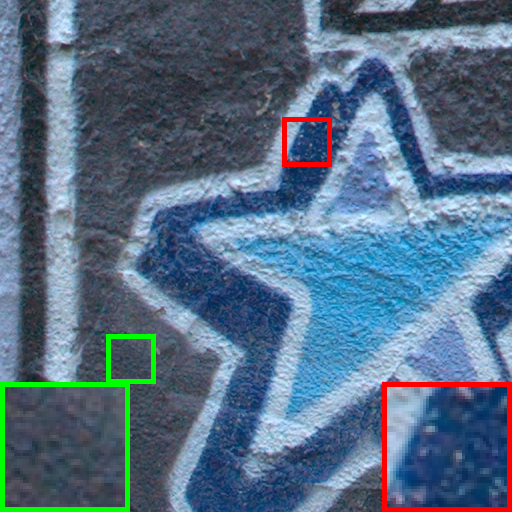}
  \centerline{\scriptsize CDnCNN-B\cite{Zhang2016Beyond}}
  \label{fig:Case2}
\end{minipage}%
}
\subfigure{
\begin{minipage}[c]{0.2\textwidth}
\centering
  \includegraphics[width=1.03\linewidth]{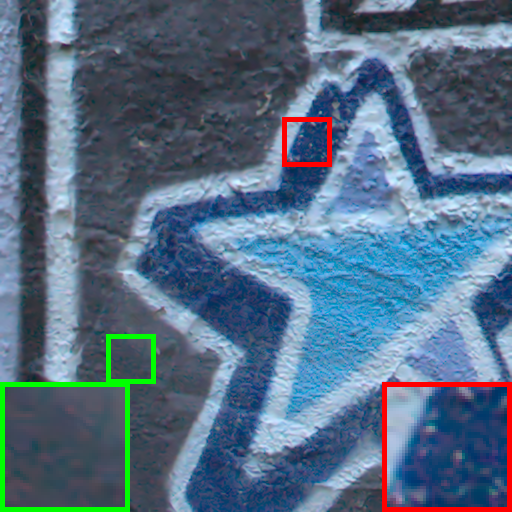}
  \centerline{\scriptsize TNRD \cite{chen2017trainable}}
  \label{fig:Case2}
\end{minipage}%
}
\subfigure{
\begin{minipage}[c]{0.2\textwidth}
\centering
  \includegraphics[width=1.03\linewidth]{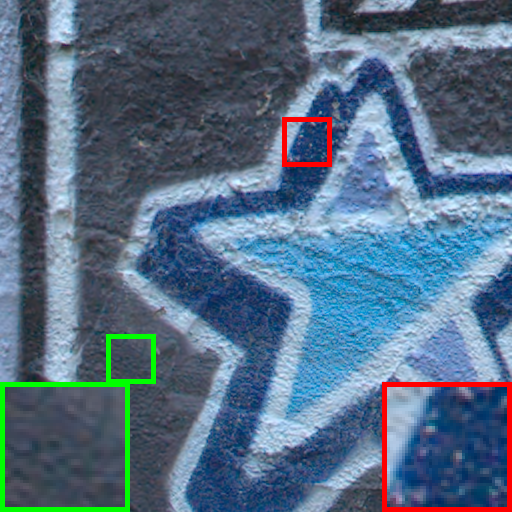}
  \centerline{\scriptsize BM3D \cite{dabov2008image}}
  \label{fig:Case2}
\end{minipage}%
}\\
\vspace{-4ex}
\subfigure{
\begin{minipage}[c]{0.2\textwidth}
\centering
  \includegraphics[width=1.03\linewidth]{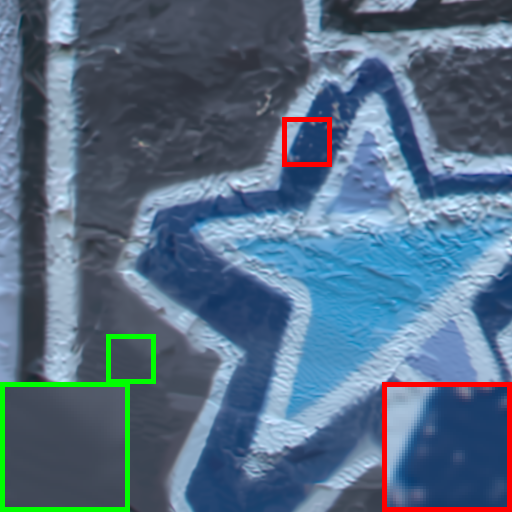}
  \centerline{\scriptsize TWSC \cite{xu2018trilateral}}
  \label{fig:Case2}
\end{minipage}%
}
\subfigure{
\begin{minipage}[c]{0.2\textwidth}
\centering
  \includegraphics[width=1.03\linewidth]{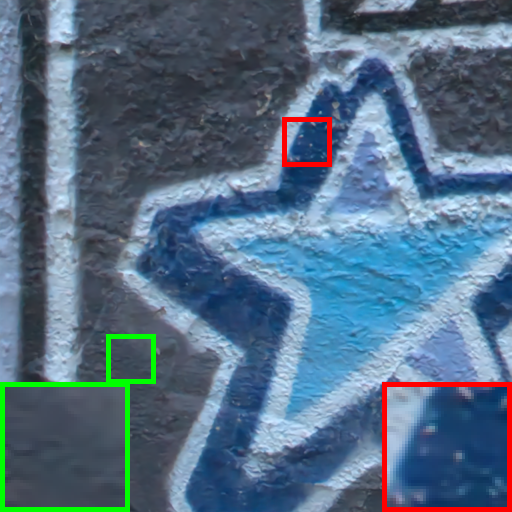}
  \centerline{\scriptsize CBDNet \cite{guo2019toward}}
  \label{fig:Case1}
\end{minipage}%
}
\subfigure{
\begin{minipage}[c]{0.2\textwidth}
\centering
  \includegraphics[width=1.03\linewidth]{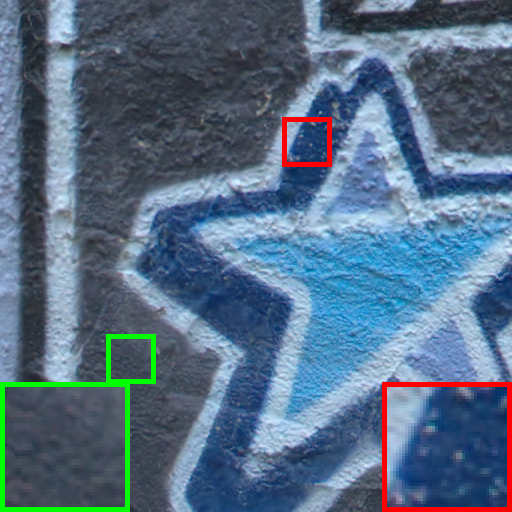}
  \centerline{\scriptsize RIDNet \cite{anwar2019real}}
  \label{fig:Case2}
\end{minipage}%
}
\subfigure{
\begin{minipage}[c]{0.2\textwidth}
\centering
  \includegraphics[width=1.03\linewidth]{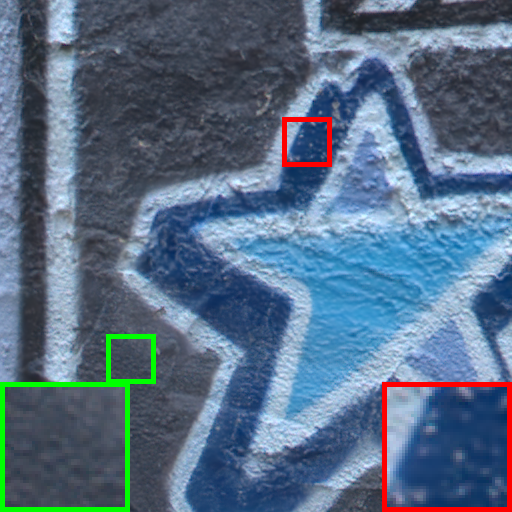}
  \centerline{\scriptsize VDN \cite{yue2019variational}}
  \label{fig:Case2}
\end{minipage}%
}\\
\vspace{-4ex}
\subfigure{
\begin{minipage}[c]{0.2\textwidth}
\centering
  \includegraphics[width=1.03\linewidth]{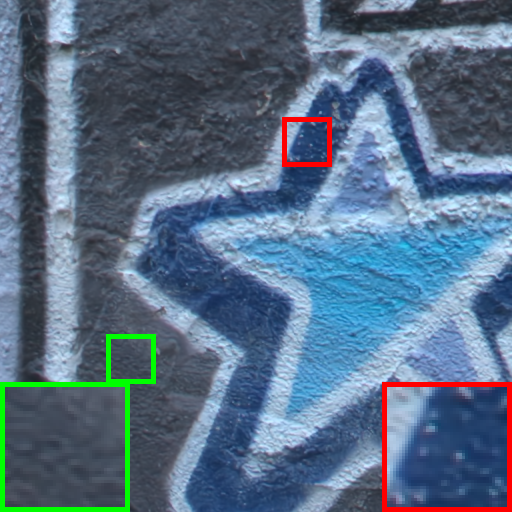}
  \centerline{\scriptsize CycleISP \cite{Zamir2020CycleISP}}
  \label{fig:Case2}
\end{minipage}%
}
\subfigure{
\begin{minipage}[c]{0.2\textwidth}
\centering
  \includegraphics[width=1.03\linewidth]{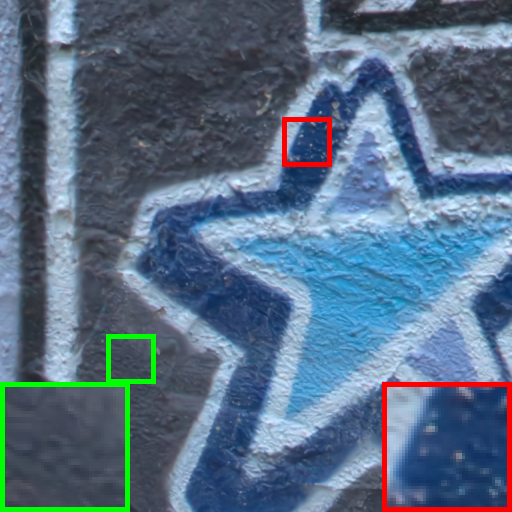}
  \centerline{\scriptsize Path-Restore \cite{yu2019path}}
  \label{fig:Case2}
\end{minipage}%
}
\subfigure{
\begin{minipage}[c]{0.2\textwidth}
\centering
  \includegraphics[width=1.03\linewidth]{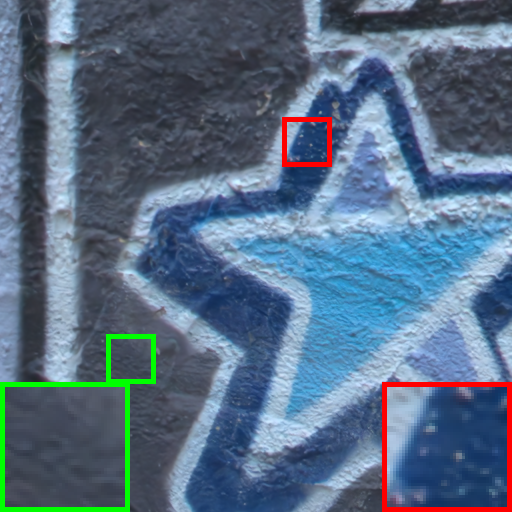}
  \centerline{\scriptsize AINDNet \cite{Kim2020Aindnet}}
  \label{fig:Case2}
\end{minipage}%
}
\subfigure{
\begin{minipage}[c]{0.2\textwidth}
\centering
  \includegraphics[width=1.03\linewidth]{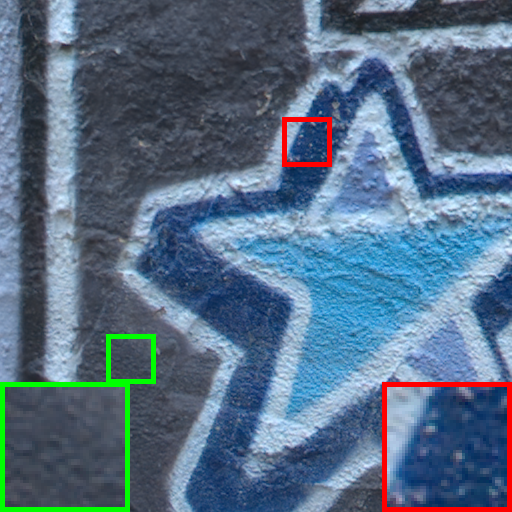}
  \centerline{\scriptsize PT-MWRN}
  \label{fig:Case2}
\end{minipage}%
}
\vspace{-1ex}
\caption{\small Denoising results of an real-world noisy image from the DND dataset by different methods.
}
\label{fig:f6}
\vspace{-4ex}
\end{center}}
\end{figure*}

Fig.~\ref{fig:f6} shows the denoising results of an image from the DND benchmark \cite{plotz2017benchmarking}.
CDnCNN-B is overfitted to AWGN removal and performs limited for real-world noisy images.
For BM3D \cite{dabov2008image} and TNRD \cite{chen2017trainable}, perceptible color artifacts can still be observed.
And over-smoothing of small-scale details and textures also cannot be avoided for the other competing methods.
%
%
%
In comparison, our PT-MWRN is more effective in preserving the color and fine-scale details faithfully.


\newcommand{\tabincell}[2]{\begin{tabular}{@{}#1@{}}#2\end{tabular}}
\begin{table}[!t]
\centering
\caption{The quantitative results on the SIDD benchmark. {\color{red}Red}, {\color{blue}{blue}} and {\textcolor[rgb]{0.03,0.69,0.94}{cyan}} {\color{black}are utilized to indicate top {\color{red}1\textsuperscript{st}}, {\color{blue}2\textsuperscript{nd}} and  {\textcolor[rgb]{0.03,0.69,0.94}{3\textsuperscript{rd}}} rank, respectively} }
\label{table}

\scalebox{0.64}{
\begin{tabular}{p{3.1cm}<{\centering}p{2.4cm}<{\centering}p{2.0cm}<{\centering}p{2.0cm}<{\centering}p{2cm}<{\centering}}
\toprule[1pt]
Method&
Blind/Non-blind&
Denoising on&
\tabincell{c}{PSNR/SSIM\\(ntire challenge)}&
\tabincell{c}{PSNR/SSIM\\(benchmark)}\\
\bottomrule[1pt]
CDnCNN-B\cite{Zhang2016Beyond}&
blind&
sRGB&
-&
23.66/0.583\\

TNRD \cite{chen2017trainable}&
Non-blind&
sRGB&
-&
24.73/0.643\\

BM3D \cite{dabov2008image}&
Non-blind&
sRGB&
-&
25.65/0.685\\

KSVD \cite{aharon2006k}&
Non-blind&
sRGB&
-&
26.88/0.842\\

CBDNet \cite{guo2019toward}&
blind&
sRGB&
-&
33.28/0.868\\

AINDNet \cite{Kim2020Aindnet}&
blind&
sRGB&
-&
39.15/0.955\\

VDN \cite{yue2019variational}&
blind&
sRGB&
-&
39.27/0.955\\

CycleISP \cite{Zamir2020CycleISP}&
blind&
sRGB&
-&
39.52/0.957\\

DIDN \cite{yu2019deep}&
blind&
sRGB&
\textcolor[rgb]{0.03,0.69,0.94}{39.818}/\textcolor[rgb]{0.03,0.69,0.94}{0.973}&
39.78/0.958\\

DHDN \cite{park2019densely}&
blind&
sRGB&
\textcolor[rgb]{0,0,1}{39.883}/\textcolor[rgb]{0,0,1}{0.9731}&
\textcolor[rgb]{0,0,1}{39.84}/\textcolor[rgb]{1,0,0}{0.959}\\

GRDN \cite{kim2019grdn}&
blind&
sRGB&
\textcolor[rgb]{1,0,0}{39.932}/\textcolor[rgb]{1,0,0}{0.9736}&
\textcolor[rgb]{1,0,0}{39.85}/\textcolor[rgb]{1,0,0}{0.959}\\

PT-MWRN(Ours)&
blind&
sRGB&
-&
\textcolor[rgb]{0.03,0.69,0.94}{39.80}/\textcolor[rgb]{1,0,0}{0.959}\\

\bottomrule[1pt]
\end{tabular}
}
\label{tab:realSIDD}
\vspace{-0.145in}
\end{table}

\textbf{SIDD.} The SIDD benchmark \cite{plotz2017benchmarking} contains 320 pairs of noisy and clean images for training, and provides 1,280 noisy images with the size $256 \times 256$ for testing.
Analogous to DND, the clean ground-truth images of the testing noisy images are not publicly available.
We compare our PT-MWRN with the state-of-the-art competitive algorithms including traditional methods BM3D \cite{dabov2008image}, KSVD \cite{aharon2006k} and deep denoisers CDnCNN-B\cite{Zhang2016Beyond}, TNRD \cite{chen2017trainable} CBDNet \cite{guo2019toward}, AINDNet \cite{Kim2020Aindnet}, VDN \cite{yue2019variational}, CycleISP \cite{Zamir2020CycleISP}, DIDN \cite{yu2019deep}, DHDN \cite{park2019densely} and GRDN \cite{kim2019grdn}.
It is worth noting that DIDN \cite{yu2019deep}, DHDN \cite{park2019densely} and GRDN \cite{kim2019grdn} are the Top 3 winner solutions of the 2019
NTIRE Real Image Denoising (RID) Challenge.

\begin{table*}[!t]\footnotesize
\centering
\caption{ Comparation with the Top 3 winner solutions of 2019 NTIRE Challenge on Real Image Denoising, i.e., GRDN \cite{kim2019grdn}, DHDN \cite{park2019densely} and DIDN \cite{yu2019deep}.}

\scalebox{1.00}{
\begin{tabular}{cccccc}

\toprule[1pt]
Method&
Time[ms]&
DND[dB]&
SIDD[dB]&
Ensemble&
Training Set\\
\bottomrule[1pt]

DIDN \cite{yu2019deep}&
221.7&
39.62&
39.78&
weight ensemble, flip/rotate($\times 8$)&
SIDD \cite{plotz2017benchmarking}\\

DHDN \cite{park2019densely}&
151.5&
39.29&
39.84&
models($\times 2$), flip/rotate($\times 8$)&
SIDD \cite{plotz2017benchmarking}\\

GRDN \cite{kim2019grdn}&
126.3&
38.70&
39.85&
None&
SIDD \cite{plotz2017benchmarking} + GAN Synthesis\\

PT-MWRN&
38.6&
39.84&
39.80&
None&
SIDD \cite{plotz2017benchmarking} + CBDNet \cite{guo2019toward} Synthesis\\

\bottomrule[1pt]

\end{tabular}
}
\label{tab:realtop}
\vspace{-0.145in}

\end{table*}

Table~\ref{tab:realSIDD} lists the PSNR/SSIM results of the competing method.
In particular, the fourth column gives the reported results of DIDN \cite{yu2019deep}, DHDN \cite{park2019densely} and GRDN \cite{kim2019grdn} on the 2019 NTIRE RID Challenge.
While the fifth column provides the denoising performance given in the official SIDD benchmark website\footnote{\url{https://www.eecs.yorku.ca/~kamel/sidd/benchmark.php}}.
The testing images from the 2019 NTIRE RID Challenge are same with those in the official SIDD benchmark website, but different data precision is adopted for evaluation, i.e., single precision (float) and integer (uint8), respectively.
So the results in the fourth and fifth columns only make little difference.
%
%
The results of GRDN \cite{kim2019grdn}, DHDN \cite{park2019densely} and DIDN \cite{yu2019deep} are based on the code and pre-trained models given in the authors' personal website.
From Table~\ref{tab:realSIDD}, it can be seen that our PT-MWRN is very competitive in comparison to the Top 3 winner solutions and performs much better than the other competing methods.

%

Finally, Table~\ref{tab:realtop} further compares our PT-MWRN with DIDN \cite{yu2019deep}, DHDN \cite{park2019densely} and GRDN \cite{kim2019grdn} on DND and SIDD in more details.
%
%
%
DHDN \cite{park2019densely} and DIDN \cite{yu2019deep} apply the self-ensemble method~\cite{lim2017enhanced} to improve denoising performance, thereby making them computationally heavy.
%
%
%
%
To illustrate this point, we also give the GPU running time (in milliseconds, ms) on a $256 \times 256$ image in the second column of Table~\ref{tab:realtop}.
%
%
In addition to SIDD dataset \cite{plotz2017benchmarking}, GRDN \cite{kim2019grdn} also applies GAN as noise simulator to generate additional synthetic training images.
Overall, our PT-MWRN is the most efficient in computation, and achieves the highest PSNR on DND and the third highest PSNR on SIDD, making it very competitive in handling real-world noisy photographs.
%


\section{Conclusion}
\label{sec:CONCLUSION}
In this paper, we presented a deep multi-level wavelet residual network (MWRN) as well as a progressive training scheme for effective image denoising.
In particular, MWRN, scale-specific loss and progressive training are incorporated to make it feasible to alleviate the difficulty of training deeper denoising networks.
For ensuring the effectiveness of scale-specific loss, the wavelet subbands of noisy image are also fed into each scale of the encoder.
%
%
%
%
Experiments show that our PT-MWRN performs favorably against the state-of-the-art methods for both Gaussian denoising on grayscale and color images, and can be easily extended for handling real-world noisy photographs with very competitive denoising performance.
Albeit our PT-MWRN is suggested for improving MWCNN for image denoising, it may also be beneficial to the training of other multi-scale CNN architectures and other image restoration tasks, which will be further studied in the future work.
\ifCLASSOPTIONcaptionsoff
  \newpage
\fi



\bibliographystyle{IEEEtran}
\bibliography{egbib}
%

%






\end{document}